\documentstyle[aps,prl,preprint,floats,epsfig,rotating]{revtex}

\begin{document}
\preprint{\tighten\vbox{\hbox{\hfil CLNS 99/1611}
                        \hbox{\hfil CLEO 99-3}
}}
% Comment out to get double spacing.
\tighten
\title{Measurement of $B \rightarrow \rho \ell \nu$ decay and 
$\vert V_{ub} \vert $} 
\date{\today}
\author{(CLEO Collaboration)}
\maketitle

\begin{abstract} 
% Insert abstract here.
Using a sample of $3.3 \times 10^6$ $\Upsilon(4S) \rightarrow
B \bar B$ events collected with the CLEO~II detector at the
Cornell Electron Storage Ring (CESR),
we measure ${\cal B}(B \rightarrow \rho \ell \nu)$,
$\vert V_{ub} \vert$, and the partial rate ($\Delta \Gamma$) in
three bins of $q^2 \equiv (p_B-p_{\rho})^2$.  We find
${\cal B}(B^0 \rightarrow \rho^- \ell^+ \nu) = (2.69 \pm 0.41^{+0.35}_{-0.40} 
\pm 0.50) \times 10^{-4}$,
$\vert V_{ub} \vert = (3.23 \pm 0.24^{+0.23}_{-0.26} \pm 0.58) 
\times 10^{-3}$,
$\Delta \Gamma (0 <q^2 < 7~{\rm GeV}^2/c^4) = 
(7.6 \pm 3.0 ^{+0.9}_{-1.2} \pm 3.0) 
\times 10^{-2} ~{\rm ns}^{-1}$,
$\Delta \Gamma (7 < q^2 < 14~{\rm GeV}^2/c^4) = 
(4.8 \pm 2.9 ^{+0.7}_{-0.8} \pm 0.7) 
\times 10^{-2} ~{\rm ns}^{-1}$, and
$\Delta \Gamma (14 < q^2 < 21~{\rm GeV}^2/c^4) = 
(7.1 \pm 2.1 ^{+0.9}_{-1.1} \pm 0.6) 
\times 10^{-2} ~{\rm ns}^{-1}$. T
Here, $\ell=e$ or $\mu$, but not both, and the 
quoted errors are statistical, systematic, and theoretical.
The method is sensitive primarily to
$B \rightarrow \rho \ell \nu$ decays with leptons in the 
energy range above 2.3~GeV.
Averaging with the previously published CLEO
results for $B \rightarrow \rho \ell \nu$, we obtain
${\cal B}(B^0 \rightarrow \rho^- \ell^+ \nu) = (2.57 \pm 0.29^{+0.33}_{-0.46}
\pm 0.41) \times 10^{-4}$ and
$\vert V_{ub} \vert = (3.25 \pm 0.14 ^{+0.21}_{-0.29} \pm 0.55) 
\times 10^{-3}$.
\end{abstract}

% Insert prd author list file here
\newpage
\begin{center}
B.~H.~Behrens,$^{1}$ W.~T.~Ford,$^{1}$ A.~Gritsan,$^{1}$
H.~Krieg,$^{1}$ J.~Roy,$^{1}$ J.~G.~Smith,$^{1}$
J.~P.~Alexander,$^{2}$ R.~Baker,$^{2}$ C.~Bebek,$^{2}$
B.~E.~Berger,$^{2}$
% ...
\end{center}
 
\small
\begin{center}
$^{1}${University of Colorado, Boulder, Colorado 80309-0390}\\
$^{2}${Cornell University, Ithaca, New York 14853}\\
% ...
\end{center}


\section{Introduction}
\label{sec:intro}

Exclusive semileptonic $b \rightarrow u \ell \nu$ decays
are an active area of experimental and theoretical 
study~\cite{lkgprl,cleoinc,alephinc,l3inc,lkgrev,wsb,ks,isgw1,isgw2,melikhov,faustov,demchuk,elc,ape,ukqcd,narison,lcsr,ruckl,e791,stech1,stech2,soares,boyd}.
These rare processes can be used to extract the magnitude of 
$V_{ub}$, one of the smallest and least well
known elements of the Cabibbo-Kobayashi-Maskawa (CKM) 
quark-mixing matrix~\cite{ckm}.  
Because $\vert V_{ub} / V_{cb} \vert \approx 0.08$,
the branching fractions for exclusive $b \rightarrow u \ell \nu$
processes are small, of order $10^{-4}$, and they have
only recently become experimentally accessible.  The present
analysis confirms the initial CLEO observation~\cite{lkgprl} of
$B \rightarrow \rho \ell \nu$, improves the precision on the
branching fraction and $\vert V_{ub} \vert$, and provides the
first information on the $q^2$ dependence for this decay.

Extracting $\vert V_{ub} \vert$ 
from a measured decay rate
requires significant theoretical input because the matrix
elements for such processes involve complex strong-interaction
dynamics.  Although the underlying
$b\to u\ell \nu$ decay is a relatively simple weak 
process, it is difficult to calculate
the strong-interaction effects involved in the
transition from the heavy $B$ meson to
the light daughter meson.
Because of these theoretical uncertainties,
even a perfectly measured $B\to\rho\ell\nu$
branching fraction would not at present lead to a precise value
of $|V_{ub}|$. 

The dynamics in $B \to \rho \ell \nu$ decay are in
contrast with
$b\to c\ell \nu$ decays, such as $B \to D^* \ell \nu$,
where a heavy quark
is present both in initial and final states. In this case,
techniques based on Heavy Quark Effective Theory (HQET) 
can be used to calculate the decay amplitude with good precision,
particularly for the kinematic configuration in which the 
charm hadron has zero recoil velocity.
The zero-recoil point in $B\to\rho\ell\nu$ cannot be
treated with similar techniques, however, because the daughter $u$ quark
is not heavy compared to the scale of hadronic energy transfers.
Nevertheless, substantial progress has
been made using a variety of theoretical methods, including
quark models~\cite{wsb,ks,isgw1,isgw2,melikhov,faustov,demchuk}, 
lattice QCD~\cite{elc,ape,ukqcd}, 
QCD sum rules~\cite{narison,lcsr,ruckl},
and models relating form factors measured in
$D \rightarrow K^* \ell \nu$ decay~\cite{e791}
to those in $B \to \rho \ell \nu$ decay. 

Experimentally, the main difficulty in
observing signals from
$b\to u\ell \nu$ processes is the very large background 
due to $b\to c\ell \nu$. Because 
a significant fraction of $B \to \rho \ell \nu$ events
have lepton energy beyond the endpoint for 
$b\to c\ell \nu$ decay, lepton-energy requirements
provide a powerful tool for background
suppression. However, extrapolation of the decay rate measured
in this portion
of phase space to the full rate again requires the
use of theoretical models, and it introduces model 
dependence beyond that associated with simply extracting
the value of $\vert V_{ub} \vert$ from the branching fraction. 
In this study, we begin to explore the decay dynamics
of $B\to\rho\ell \nu$ by measuring the distribution
of $q^2$, the square of the mass of the virtual $W$. The
distribution of $q^2$ is reflected in the $\rho$ momentum
spectrum.  Eventually,
studies of the $q^2$ distribution, as well as of the 
angular distributions of the decay products,
should reduce the model dependence on
$\vert V_{ub} \vert$ by constraining theoretical models
for the decay form factors.

CLEO has previously observed~\cite{lkgprl} the decays
$B^0 \rightarrow \rho^- \ell^+ \nu$, $B^+ \rightarrow \rho^0 \ell^+ \nu$,
$B^0 \rightarrow \pi^- \ell^+ \nu$, and
$B^+ \rightarrow \pi^0 \ell^+ \nu$
by measuring both the missing
energy and momentum in an event to infer the momentum of the
neutrino.  Using $2.84 \times 10^6$ $\Upsilon(4S) \to B \bar B$
events, this study obtained
\begin{eqnarray}
{\cal B}(B^0 \rightarrow \rho^- \ell^+ \nu) &=& ( 2.5 \pm 0.4^{+0.5}_{-0.7}
\pm 0.5 ) \times 10^{-4} ,\nonumber \\
{\cal B}(B^0 \rightarrow \pi^- \ell^+ \nu) &=& ( 1.8 \pm 0.4 \pm 0.3
\pm 0.2 ) \times 10^{-4}, \nonumber \\
\vert V_{ub} \vert &=& (3.3 \pm 0.2^{+0.3}_{-0.4} \pm 0.7) \times 10^{-3},
\label{eq:lkgnums}
\end{eqnarray}
\noindent
where the errors are statistical, systematic, and theoretical.
A large contribution to the $B \rightarrow \rho \ell \nu$
systematic error is associated with a possible nonresonant 
$B \rightarrow \pi \pi \ell \nu$ rate.  In the analysis described
below this uncertainty is reduced.

We report on a measurement of 
${\cal B}(B \rightarrow \rho \ell \nu)$ and
$\vert V_{ub} \vert$ as well as the first measurement of
the partial rate for $B \rightarrow \rho \ell \nu$
in bins of $q^2$.  We study five signal modes:
$B^0 \to \rho^- \ell^+ \nu$, $B^+ \to \rho^0 \ell^+ \nu$,
$B^+ \to \omega \ell^+ \nu$, $B^0 \to \pi^- \ell^+ \nu$,
and $B^+ \to \pi^0 \ell^+ \nu$.
Our method is sensitive primarily
to leptons with energies above the
$b \rightarrow c \ell \nu$ lepton-energy 
endpoint ($E_{\ell} > 2.3~{\rm GeV}$).  
The resulting event sample is essentially statistically
independent of that from the previous analysis
since it contains much larger signal and background
yields, even though the data samples are similar.

In this paper, we use the notation $b \to u (c) \ell \nu$
to denote inclusive $B \to X_{u(c)} \ell \nu$ decay,
where $X_u$ is a hadronic system with no charm quarks, and
$X_c$ is a hadronic system with a charm quark.  When
discussing backgrounds, we will often refer to $b \to u \ell \nu$
inclusive decays excluding one or more
signal modes.  In this case, we will use the notation
$b \to u \ell \nu$ and explicitly list the modes not to be
included.  When no charge state is given,
$B \to \rho \ell \nu$ refers generically to 
$B \to \rho^{\pm} \ell \nu$,
and $B \to \rho^0 \ell \nu$ decays.
When specifying a particular decay mode, we implicitly
include its charge-conjugate decay.

We begin in Sec.~\ref{sec:kine}
by introducing the phenomenology of $B \rightarrow \rho \ell 
\nu$ decay.  The full differential decay rate is expressed in terms
of helicity amplitudes, which depend in turn on form factors,
Lorentz-invariant functions of $q^2$ that parametrize
the hadronic current.
We discuss five form-factor models that are used to obtain
$\vert V_{ub} \vert$ and
to extrapolate our yield measured at high
lepton energy to the 
full phase space available in $B \rightarrow \rho \ell \nu$
decay.  

Section~\ref{sec:events} describes the data sample and
the requirements used to distinguish
signal events from backgrounds, which are due primarily to 
continuum events ($e^+ e^- \to q \bar q$, $q \bar q 
= u \bar u$, $d \bar d$, $s \bar s$, and
$c \bar c$),
$ b \rightarrow c \ell \nu$,
and $b \rightarrow u \ell \nu$ (other than
$B \to \rho \ell \nu$ or $B \to \omega \ell \nu$) events.

Section~\ref{sec:anal} describes the 
binned maximum-likelihood fit used to extract the 
$B \rightarrow \rho \ell \nu$ signal.
Although we must
rely on Monte Carlo calculations to model the shapes of distributions for the
$b \rightarrow c(u) \ell \nu$ backgrounds, the
normalization and lepton-energy spectrum 
of each background component are determined by the data.
In addition,
the assumed background shapes are extensively tested
using sideband regions where the contribution from signal
events is small.  The continuum background contribution is
measured directly using data.
Section~\ref{sec:fitresults} presents the 
yields extracted from the fit and the
model-dependent detection efficiencies used to obtain values
for ${\cal B}(B \rightarrow \rho \ell \nu)$
and $\vert V_{ub} \vert$.  

Section~\ref{sec:sys} describes the contributions to our
systematic error.
Apart from model dependence, these  
are due primarily to uncertainties in the $b \rightarrow u \ell \nu$ 
backgrounds and to uncertainties in the
detector simulation.  In the high lepton-energy region,
where the sensitivity to the signal is greatest, the
background from other $b \to u \ell \nu$ processes
is comparable to that from $b \to c \ell \nu$ decays, but
its kinematic properties are less well understood.

Section~\ref{sec:res} presents our measured values of
${\cal B}(B \to \rho \ell \nu)$ and $\vert V_{ub} \vert$,
as well as the average with the 
previous CLEO result. 
This average takes into account
the correlation in the
systematic errors as well as
the variation in results for different form-factor models.
In Sec.~\ref{sec:conclusion} we present our conclusions
and an outlook for future measurements.
\section{Semileptonic decay kinematics}
\label{sec:kine}

The matrix element for the semileptonic decay 
$P(Q \bar q) \rightarrow V(q^{\prime} \bar q) \ell^- \bar \nu$
of a pseudoscalar meson, $P(Q \bar q)$, to a vector meson, 
$V(q^{\prime} \bar q)$, 
can be written~\cite{richman}
\begin{equation}
%equation 9 in richman+Burchat
{\cal M}(P \rightarrow V \ell \nu) = -{\it i} {{G_F} \over
{\sqrt{2}}} V_{q^{\prime}Q} L^{\mu} J_{\mu},
\end{equation}
where $V_{q^{\prime}Q}$ is the CKM matrix element for the
$Q \to q^{\prime}$ transition~\cite{vqq},
$L^{\mu}$ is the leptonic current
\begin{equation}
%equation 10 in richman+Burchat
L^{\mu} = \bar u_{\ell} \gamma^{\mu} (1 - \gamma_5) v_{\nu},
\end{equation}
and the hadronic current
\begin{eqnarray}
%equation 85 in richman+Burchat
J_{\mu} = \left < V(p_V, \varepsilon) \vert (V-A)_{\mu} \vert P(p_P) 
\right > = 
 &-&\varepsilon^*_{\mu} (m_P+m_V) A_1(q^2) \nonumber \\
&+& (p_V+p_P)_{\mu} ( \varepsilon^* \cdot q) 
{ { A_2(q^2)} \over {(m_P + m_V)}} 
\nonumber \\
&+& q_{\mu} ( \varepsilon^* \cdot q) { {2 m_V} \over {q^2}} \bigl (A_3(q^2)
- A_0(q^2) \bigr ) \nonumber \\ &+&
2i\epsilon_{\mu \nu \rho \sigma} \varepsilon^{* \nu} p_V^{\rho} p_P^{\sigma}
{ { { V}(q^2)} \over { m_P + m_V} },
\end{eqnarray}
is written in
terms of four form factors, $A_1(q^2)$, $A_2(q^2)$,
$V(q^2)$, and $A_0(q^2)$,
where $A_0 (0) = A_3(0)$ and
\begin{equation}
%equation 86 in richman+Burchat
A_3(q^2) = {{m_P+ m_V} \over {2 m_V}} A_1(q^2) -
{{m_P- m_V} \over {2 m_V}} A_2(q^2).
\end{equation}
\noindent
Here $m_P$, $m_V$, $p_{P}$, and $p_{V}$ are the 
pseudoscalar and vector
meson masses and four-momenta, $q \equiv p_{P} - p_{V}$, and
$\varepsilon$ is the vector-meson polarization four-vector.
Terms in $J_{\mu}$ 
proportional to $q_{\mu}$ vanish in the limit of massless
leptons, so that the decay $B \rightarrow \rho \ell \nu$ depends
effectively on only
three form factors ($A_1(q^2)$, $A_2(q^2)$, and $V(q^2)$)
for electrons or muons.  

The amplitudes for the vector meson to have helicity $+1$, $-1$, or $0$,
denoted by $H_+(q^2)$, $H_-(q^2)$, and
$H_0(q^2)$, can be expressed in terms of the form factors:
\begin{eqnarray}
%equation 116 in richman+Burchat
H_{\pm}(q^2) &=& (m_P + m_V)A_1(q^2) \mp {{2m_P \vert {\bf p}_V \vert} \over 
{m_P+m_V}} V(q^2), \nonumber \\
%equation 115 in richman+Burchat
H_0(q^2) &=& {{1} \over {2 m_V \sqrt{q^2}}} \bigl [ (m_P^2 - m_V^2 - q^2)
(m_P+m_V)A_1(q^2) \nonumber \\
&-&4 {{m_P^2 \vert {\bf p}_V \vert^2} \over {m_P+m_V}} A_2(q^2) \bigr ],
\end{eqnarray}

\noindent where ${\bf p}_V$ is the vector meson three-momentum
in the $B$-meson rest frame, which is a function of $q^2$
\begin{equation}
\vert {\bf p}_V \vert = \sqrt{ \left(
{ {m_P^2 + m_V^2 - q^2} \over {2 m_P}} \right)^2 - m_V^2}.
\end{equation}
As $q^2$ decreases from $q^2_{{\rm max}} = (m_P - m_V)^2$
to $q^2_{{\rm min}} = m_{\ell}^2 \approx 0$, 
$\vert {\bf p}_V \vert$ increases from 
$\vert {\bf p}_V \vert_{{\rm min}}=0$ to
$\vert {\bf p}_V \vert_{{\rm max}}=(m_P^2-m_V^2)/2m_P^2$.
At large $q^2$, the $A_1(q^2)$ form factor therefore dominates all
three of the helicity amplitudes. 
The full differential decay rate for $P \rightarrow V \ell \nu$
($V \rightarrow P_1 P_2$) is given by~\cite{richman}
\begin{eqnarray}
%equation 113 in richman+Burchat
{{d \Gamma} \over {dq^2 d \cos \theta_{\ell} d \cos \theta_{V} d \chi}} &=& 
{{3} \over {8 (4 \pi)^4}} G_F^2 \vert V_{q^{\prime}Q} \vert^2 {
{\vert {\bf p_{V}} \vert  q^2}
\over {m_P^2}} {\cal B}(V \rightarrow P_1 P_2) \nonumber\\
&\times& \biggl [
(1- \eta \cos \theta_{\ell})^2 \sin^2 \theta_V \vert H_+(q^2) \vert^2 
\nonumber \\
&+&(1+ \eta
\cos \theta_{\ell})^2 \sin^2 \theta_V \vert H_-(q^2) \vert^2 \nonumber \\ 
&+&4 \sin^2 \theta_{\ell} \cos^2 \theta_{V} \vert H_0(q^2) \vert^2 \nonumber \\
&-&4 \eta \sin \theta_{\ell} (1- \eta \cos \theta_{\ell}) \sin \theta_{V}
\cos \theta_{V} \cos \chi H_+(q^2) H_0(q^2)  \nonumber \\ 
&+&4 \eta \sin \theta_{\ell} (1+\eta \cos \theta_{\ell}) \sin \theta_{V}
\cos \theta_{V} \cos \chi H_-(q^2) H_0(q^2) \nonumber \\ 
&-&2 \sin^2 \theta_{\ell} \sin^2 \theta_V \cos {2 \chi} H_+(q^2) H_-(q^2)
\biggr ],
\label{eq:4d} 
\end{eqnarray}
where $q^2 = (p_P - p_{V})^2$, $\theta_{\ell}$ is 
the polar angle of the lepton in the $W$ rest frame
with respect to the $W$ 
flight direction in the $B$ rest frame, and $\theta_{V}$ is 
the polar angle of one of the pseudoscalar daughters
in the rest frame of the vector meson
with respect to the vector-meson 
flight direction in the $B$ rest frame.  $\chi$ is
the angle between the decay planes of the $W$ and the vector
meson. The factor $\eta$ is equal to $+1~(-1)$ when the quark $Q$ has
charge $-1/3$ ($+2/3$).

\begin{figure}
\begin{center}
\begin{turn}{-90}
\epsfig{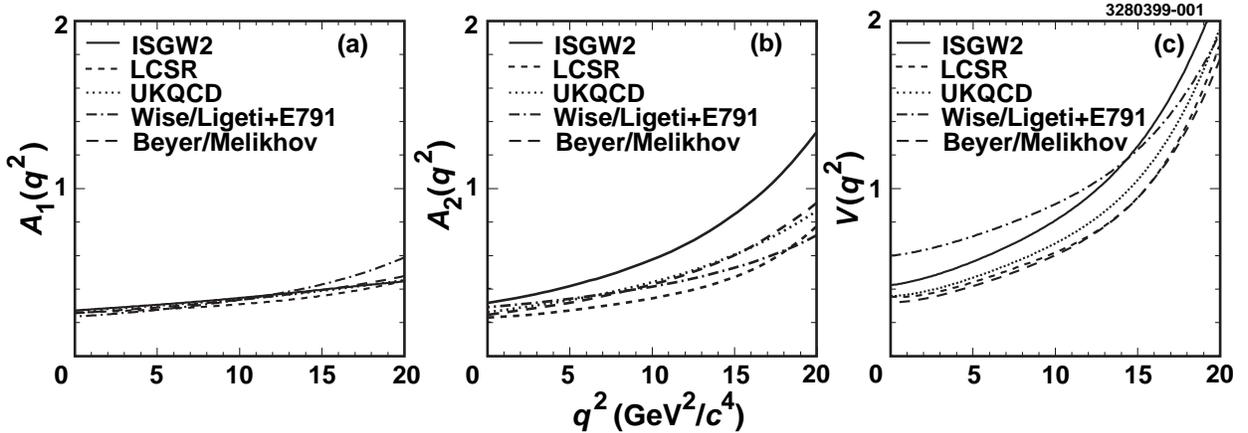}
\end{turn}
\end{center}
\caption{Model predictions for the (a) $A_1(q^2)$, 
(b) $A_2(q^2)$, and (c) $V(q^2)$ 
form factors.  The $A_1(q^2)$ form factor determines the
rate at high $q^2$.  The $V(q^2)$ form factor has a large
effect on the lepton-energy spectrum.  Maximum daughter-hadron
recoil occurs at the minimum value of $q^2$, which is close to zero.}
\label{fig:formfactors}
\end{figure}

For the extraction
of ${\cal B}(B \rightarrow \rho \ell \nu)$ and $\Delta \Gamma$
in bins of $q^2$
from the data, we use Eq.~\ref{eq:4d}
to extrapolate
our measurements based primarily on the high lepton energy
part of the decay phase space to the entire phase space.
The form factors $A_1(q^2)$, $A_2(q^2)$, and $V(q^2)$ are
evaluated using various models, whose predictions are
shown in Fig.~\ref{fig:formfactors}.  
Models additionally provide the
normalization $\tilde{\Gamma}_{{\rm thy}} \equiv \Gamma(B
\rightarrow \rho \ell \nu) / \vert V_{ub} \vert^2$, which
allows us to extract $\vert V_{ub} \vert$ from the measured
decay rate.  We
use a set of form-factor models representative
of current theoretical work, including results from
two quark-model approaches (ISGW2~\cite{isgw2} 
and Melikhov/Beyer~\cite{melikhov}), 
lattice QCD (UKQCD~\cite{ukqcd}), 
light-cone sum rules (LCSR~\cite{lcsr}), and a method incorporating
form-factor measurements
in $D \rightarrow K^* \ell \nu$ decay (Wise/Ligeti+E791~\cite{e791}).
In all
cases we have used form-factor parametrizations as 
described in the above references.
A theoretical error is added to the branching fraction
and $\Delta \Gamma$
measurements based on
one-half of the full
spread in results among these models.  The $\vert V_{ub} \vert$
measurement has an additional theoretical uncertainty due
to the determination of $\tilde{\Gamma}_{{\rm thy}}$.

%%%%%%%%%%%Put figure 1 here
%%%%%%%%%%%Put figure 1 here

\begin{figure}
\begin{center}
\epsfig{figure=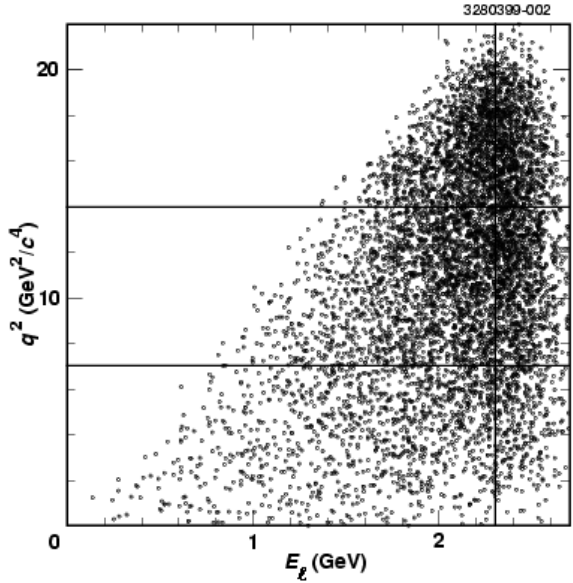,width=3.0in}
\end{center}
\caption{A simulated Dalitz plot, $q^2$ vs.~$E_{\ell}$,
for $B \to \rho \ell \nu$ decay.  
The lepton energy, $E_{\ell}$, is computed in the $\Upsilon(4S)
\to B \bar B$ rest frame (lab frame), which 
differs slightly from the $B$ rest frame.
This analysis is most sensitive to $B \to \rho \ell \nu$ events in
the lepton-energy region to the right of the vertical 
line, $E_{\ell} > 2.3$~GeV.
The horizontal lines define the $q^2$ bins in the
$\Delta \Gamma$ measurement: $0<q^2<7~ {\rm GeV}^2/c^4$,
$7<q^2<14~ {\rm GeV}^2/c^4$, and $14<q^2<21~ {\rm GeV}^2/c^4$.
At high $q^2$, our lepton-energy
requirement retains a much larger fraction of the phase
space than it does at low $q^2$.}
\label{fig:dalitz}
\end{figure}

The Dalitz plot for $B \to \rho \ell \nu$
decay predicted by the ISGW2 model is shown in Fig.~\ref{fig:dalitz}.
For a given value of $q^2$, $\cos \theta_{\ell}$ varies
from $-1$ to $+1$ across the allowed range of lepton energies.  
The Dalitz plot shows that positive values of $\cos \theta_{\ell}$
are favored over negative values; this effect, which produces
a hard lepton-energy spectrum, is a consequence of the
$V-A$ coupling, and is discussed further below.
We are most sensitive to $B \to \rho \ell \nu$ decay in 
the endpoint region to the right of the vertical line, $E_{\ell} > 2.3$~GeV.
Although a significant
fraction of the total predicted rate for $B \rightarrow \rho \ell \nu$
lies at high lepton energy, Table~\ref{table:q2info} shows that
theoretical approaches differ, giving results
between 24$\%$ and 35$\%$ for this fraction.  

Table~\ref{table:q2info} also shows the quantity
$\tilde{\Gamma}_{{\rm thy}}$
predicted by each form-factor model.  
Even with a perfectly measured $B \to \rho \ell \nu$
branching fraction, $\vert V_{ub} \vert$ would be
uncertain due to the spread in $\tilde{\Gamma}_{{\rm thy}}$
values among form-factor models.  We can summarize
the model predictions
as
$\tilde{\Gamma}_{{\rm thy}} = (16.8 \pm 2.6) ~{\rm ps}^{-1}$,
a $15 \%$ spread.  As explained in Section~\ref{sec:res},
we assign a $30 \%$ error on $\tilde{\Gamma}_{{\rm thy}}$.

%%%%%%%%%%Put figure 1 here

\begin{figure}
\begin{center}
\begin{turn}{-90}
\epsfig{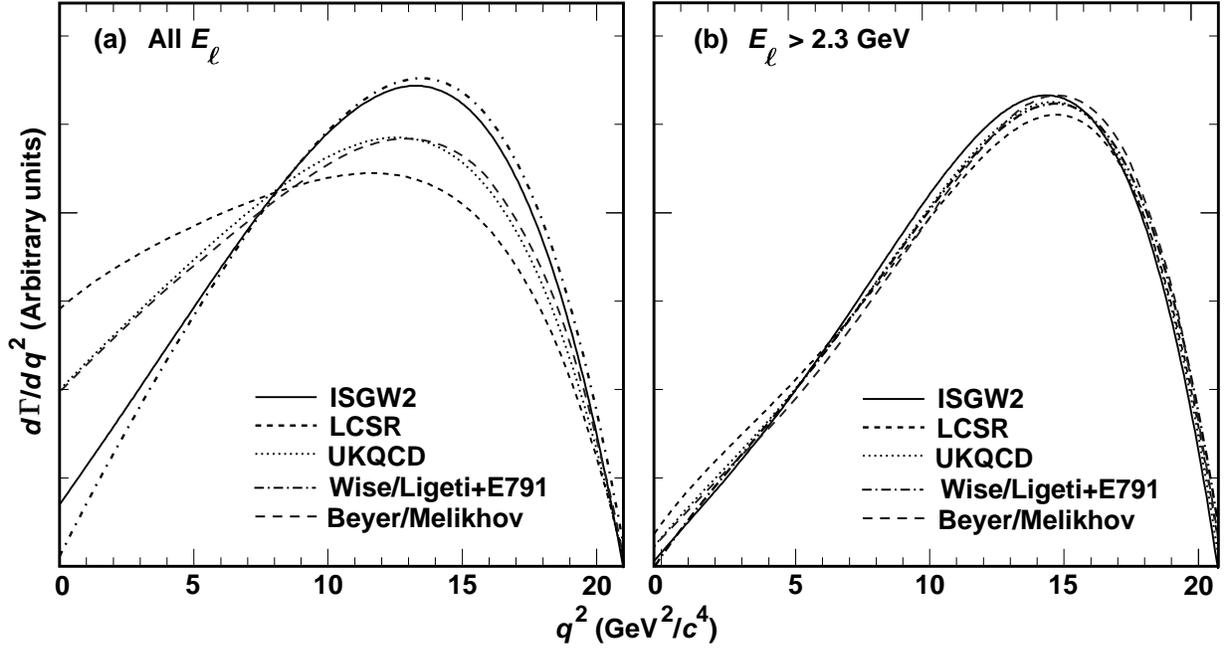}
\end{turn}
\end{center}
\caption{Theoretical $q^2$ distributions for the 
$B \rightarrow \rho \ell \nu$ form-factor models that we have
considered: (a) over the full lepton-energy
range and (b) for $E_{\ell} > 2.3$~GeV. 
The lepton energy, $E_{\ell}$, is computed in the $\Upsilon(4S)
\to B \bar B$ rest frame (lab frame), which 
differs slightly from the $B$ rest frame.
In each plot, models have been normalized to equal areas.
For high lepton energies, $E_{\ell}>2.3$~GeV, all of the models
predict very similar $q^2$ distributions.}
\label{fig:q2model}
\end{figure}
%%%%%%%%%%Put figure 4 here

\begin{table}
\begin{center}
\begin{tabular}{cccc}
FF model & $\tilde{\Gamma}_{{\rm thy}}$ (ps$^{-1}$) & 
$\Gamma(E_{\ell} > 2.3~{\rm GeV})/\Gamma$ ($\%$) &
$\Gamma(2.0 < E_{\ell} < 2.3~{\rm GeV})/\Gamma$ ($\%$) \\ \hline
ISGW2            &14.2 & $35 $ & $33 $ \\
LCSR             &16.9 & $24 $ & $28 $ \\
UKQCD            &16.5 & $27 $ & $30 $ \\
Wise/Ligeti+E791 &19.4 & $31 $ & $34 $ \\
Beyer/Melikhov   &16.0 & $27 $ & $30 $ \\
\end{tabular}
\end{center}
\caption{$\tilde{\Gamma}_{{\rm thy}}$ and lepton-energy distribution
predictions from each form-factor (FF) model.  
The lepton energy, $E_{\ell}$, is computed in the $\Upsilon(4S)
\to B \bar B$ rest frame (lab frame), which 
differs slightly from the $B$ rest frame.
For the branching fraction measurement,
the primary source of model dependence is in the ratio
$\Gamma(E_{\ell} > 2.3~{\rm GeV})/\Gamma$.  For the $\vert V_{ub} \vert$
measurement, there is significant additional model 
dependence arising from 
the quantity $\tilde{\Gamma}_{{\rm thy}}$.}
\label{table:q2info}
\end{table}

Figure~\ref{fig:q2model} shows the $q^2$ distribution
predicted by our implementation of these models.
Integrated over all lepton energies, there is a substantial
difference among the models at low $q^2$ (large daughter-hadron
recoil).
At high lepton energy, however,
the shapes of the $q^2$ distributions predicted by each of the
form-factor models are very similar.  To 
distinguish among form-factor models on the basis of the
$q^2$ distribution, the lepton-energy
region below 2.0~GeV must be probed.  

Table~\ref{table:elepeffvsq2} shows how the 
efficiency of the lepton-energy requirement varies with
$q^2$ for each form-factor model.
Our lepton-energy
requirement retains a much larger fraction of the phase
space at high $q^2$ than it does at low $q^2$. Thus,
at high $q^2$, the efficiency is quite high and a small variation
is seen among models.  At low $q^2$, on the other
hand, the efficiency is relatively low with a large
variation among the form-factor models.

Integrating Eq.~\ref{eq:4d} 
over the angular variables, we obtain
\begin{equation}
% ryd equation 2.79
{{d \Gamma} \over {d q^2}} = 
{{G_F^2} \over {96 \pi^3}}  \vert V_{q^{\prime}Q} \vert^2 {{
\vert {\bf p}_{V} \vert q^2}
\over {m_P^2}} {\cal B}(V \rightarrow P_1 P_2) 
\times \biggl [ \vert H_+ (q^2) \vert^2 + 
\vert H_- (q^2) \vert^2 + \vert H_0 (q^2) \vert^2 \biggr ].
\label{eq:dgdq}
\end{equation}

\begin{figure}
\begin{center}
\begin{turn}{-90}
\epsfig{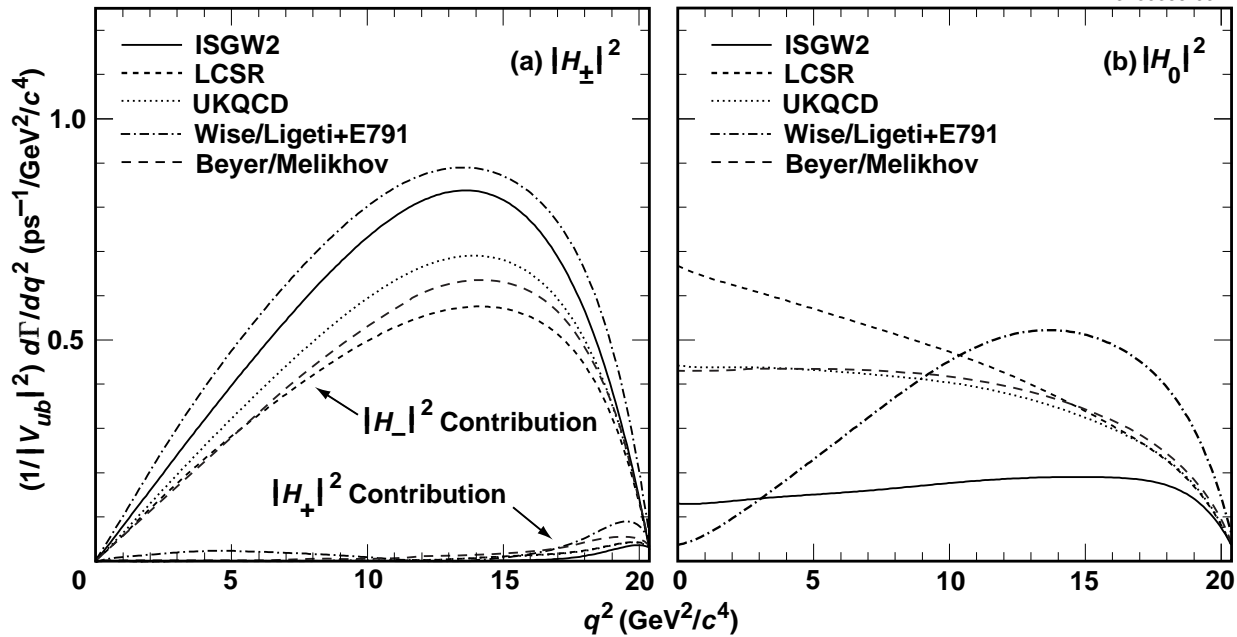}
\end{turn}
\end{center}
\caption{$d \Gamma / d q^2$ distributions for each of the
three terms in Eq.~\ref{eq:dgdq}: (a) 
the terms proportional to $\vert H_- \vert^2$ and
$\vert H_+ \vert^2$ and (b) the 
$\vert H_0 \vert^2$ term.}
\label{fig:helamps}
\end{figure}

\noindent Figure~\ref{fig:helamps} shows the $d \Gamma / d q^2$ distribution
for each term in Eq.~\ref{eq:dgdq}.  In each form-factor
model, the term proportional to $\vert H_- \vert^2$ contributes
the largest fraction of the total rate.  As a 
consequence of the $V-A$ couplings of
the $W$ boson the vector meson is more likely to have
helicity $-1$ than $+1$.  For $B \to \rho \ell \nu$ decay, this
asymmetry is quite large, except at very high $q^2$ (low 
daughter-hadron recoil momentum) where the $A_1$ form factor dominates
each of the helicity amplitudes and the $\rho$ is
nearly unpolarized, and at small values of
$q^2$, where only the zero-helicity component
contributes significantly.  Because the 
$\vert H_+ \vert^2$ contribution is quite small, the theoretical
error in the ${\cal B}(B \to \rho \ell \nu)$
and $\Delta \Gamma$ measurements
is due primarily to the uncertainty in
the relative size of  
$H_-(q^2)$ and
$H_0(q^2)$.  

Table~\ref{table:gammaq2bin} shows the predictions for
$\Gamma / \vert V_{ub} \vert^2$ in  
bins of $q^2$ ($\Delta \tilde{\Gamma}_{{\rm thy}}$) 
for the form-factor models.  These predictions can
be used to extract $\vert V_{ub} \vert$ from a 
measurement of the $q^2$ distribution.

%%%%%%%%%%%Put figure 1 here

\begin{table}
\begin{center}
\begin{tabular}{cccc}
FF model & $\Delta \Gamma(E_{\ell} > 2.3~{\rm GeV})/ \Delta \Gamma$ 
& $\Delta \Gamma(E_{\ell} > 2.3~{\rm GeV})/\Delta \Gamma$ 
& $\Delta \Gamma(E_{\ell} > 2.3~{\rm GeV})/\Delta \Gamma$  \\
 & ($\%$) & ($\%$) & ($\%$) \\
  & $0<q^2<7$~GeV$^2/c^4$ &
 $7<q^2<14$~GeV$^2/c^4$ &
 $14<q^2<21$~GeV$^2/c^4$ \\ \hline
ISGW2            & $24 $ & $35 $ & $43 $\\
LCSR             & $12 $ & $25 $ & $38 $\\
UKQCD            & $15 $ & $28 $ & $39 $\\
Wise/Ligeti+E791 & $23 $ & $29 $ & $37 $\\
Beyer/Melikhov   & $14 $ & $27 $ & $38 $  \\ 
\end{tabular}
\end{center}
\caption{The fraction of events predicted to
have $E_{\ell} > 2.3$~GeV in bins of $q^2$ for each form-factor (FF)
model. The spread among these results
is the largest component of
the theoretical error in the $\Delta \Gamma$ 
measurement.}
\label{table:elepeffvsq2}
\end{table}

\begin{table}
\begin{center}
\begin{tabular}{cccc}
FF model & $\Delta \tilde{\Gamma}_{{\rm thy}}$ (ps$^{-1}$) & 
 $\Delta \tilde{\Gamma}_{{\rm thy}}$ (ps$^{-1}$) & 
 $\Delta \tilde{\Gamma}_{{\rm thy}}$ (ps$^{-1}$) \\ 
  & $0<q^2<7$~GeV$^2/c^4$ &
 $7<q^2<14$~GeV$^2/c^4$ &
 $14<q^2<21$~GeV$^2/c^4$ \\ \hline
ISGW2            &2.8 & 6.1 & 5.3 \\
LCSR             &5.4 & 6.6 & 4.9 \\
UKQCD            &4.5 & 6.7 & 5.3 \\
Wise/Ligeti+E791 &3.5 & 8.4 & 7.5 \\
Beyer/Melikhov   &4.3 & 6.4 & 5.3 \\
\end{tabular}
\end{center}
\caption{Partial rate
($\Delta \tilde{\Gamma}_{{\rm thy}}$) predictions in $q^2$ bins
from each form-factor (FF) model. } 
\label{table:gammaq2bin}
\end{table}

\section{Data sample and event selection}
\label{sec:events}

The data used in this analysis were collected using the 
CLEO~II detector~\cite{cleoii} located at the 
Cornell Electron Storage Ring (CESR), operating
near the $\Upsilon(4S)$ resonance.  We
have analysed a $3.1$~fb$^{-1}$ sample 
taken on the resonance,
corresponding to approximately $3.3 \times 10^6$
$B \bar B$ pairs.  Additionally, we examine
a $1.7$~fb$^{-1}$ sample taken at a center-of-mass
energy $60$~MeV below the $\Upsilon(4S)$ mass.  These
off-resonance data, which have an energy below
the threshold for $B \bar B$ production, are used 
to measure the continuum background. 

The CLEO II detector is 
designed to provide excellent charged- and neutral-particle 
reconstruction efficiency and resolution.  
Three cylindrical tracking chambers are surrounded by
a time-of-flight (TOF) system, a CsI calorimeter, and
a muon tracking system.  The nearly complete
solid-angle coverage of both the tracking system and
calorimeter is exploited in this analysis to obtain information
on the momentum of the neutrino.

To be considered as a
hadron or lepton candidate, charged tracks must
satisfy several requirements.  The impact parameter
of the track 
along and transverse to 
the beam direction must be less than 5~cm and 2~mm, respectively. 
The spread in the 
impact parameter along the beam direction is dominated
by the beam width in that direction and
is approximately 1.5~cm.
Transverse to the beam direction, the resolution on
the impact parameter is approximately $0.3$~mm.
The r.m.s. residual for the
hits on the track must be less than 1~mm.  
Finally, we require that 
at least 15 out of the 67 tracking layers 
be used in the track fit.

Electron and muon identification requirements are chosen
to achieve high efficiency 
and a low hadronic fake rate.
We require $\vert \cos \theta \vert < 0.85$, where $\theta$
is the polar angle of the track's momentum vector
with respect to the beam axis.
Electrons are identified primarily using the ratio 
of the calorimeter energy
to track momentum ($E/p$) and specific ionization
($dE/dx$) information.  
Muon candidates are found by extrapolating tracks from
the central detector into the muon counters; such   
candidates are required to penetrate at least seven interaction
lengths of iron.
For electrons (muons) that satisfy the tracking requirements
described above as well as the lepton-energy
requirements of our analysis, the efficiency of these
lepton-identification requirements is approximately $94 \%$
($82 \%$).
The fraction of hadrons passing electron (muon) identification
criteria is determined from data and is 
approximately $0.1\%~(0.6 \%)$. 

Photon candidates are associated with CsI calorimeter clusters
that are not matched to any charged track.
Individual photons are required to have energy, $E_{\gamma}$,
greater than 30~MeV.
To be considered in a 
$\pi^0$ candidate, two photons must
have a combined invariant mass,
$M_{\gamma \gamma}$, within
$2\sigma$ of the nominal $\pi^0$ mass.  Their combined
energy, $E_{\gamma \gamma}$, is required to be greater than
325~MeV.  In addition, we require that at least one photon 
satisfy $\vert \cos \theta_{\gamma} \vert < 0.71$, where
$\theta_{\gamma}$ is the
polar angle with respect to the beam axis.
Each photon candidate is
included in at most one $\pi^0$ candidate.

To search for $B \rightarrow \rho \ell \nu$,
$B \rightarrow \omega \ell \nu$, and $B \rightarrow \pi \ell \nu$
events,
electron or muon candidates are combined with 
$\pi^{\pm} \pi^0$, $\pi^+ \pi^-$, $\pi^+ \pi^- \pi^0$, $\pi^{\pm}$,
or $\pi^0$ candidates.  Combinatoric backgrounds are suppressed
by requiring candidates for these hadronic systems 
to have a momentum greater than 300~MeV/$c$;  these 
backgrounds are further suppressed in the
$\pi^+ \pi^- \pi^0$ mode by making a cut on
the $\omega \to \pi^+ \pi^- \pi^0$ 
Dalitz amplitude.  We require that the
$\pi^+$, $\pi^-$, and $\pi^0$ be configured such that
the decay probability density ($d \Gamma \propto \vert {\bf p}_{\pi^+}
\times {\bf p}_{\pi^-} \vert^2$ where
${\bf p}_{\pi^+}$ and ${\bf p}_{\pi^-}$ are evaluated in
the $\omega$ rest frame)
is at least $12 \%$ of its
maximum value.

We define the missing momentum in the event to be
\begin{equation}
{\bf p}_{{\rm miss}} \equiv - \Sigma {\bf p}_i,
\label{eq:pmiss}
\end{equation}
where the sum is over reconstructed charged tracks
and photon candidates in the event.  If the event contains 
no undetected particles other than the $\nu$ from the
$B \to \rho \ell \nu$ decay,
${\bf p}_{{\rm miss}} \approx {\bf p}_{{\nu}}$. 
The resolution on ${\bf p}_{{\rm miss}}$ is
determined by undetected particles, such as
$K^0_L$ mesons, and our experimental resolution.  As
discussed in the following section, 
${\bf p}_{{\rm miss}}$ is used to calculate
one of our three fit variables.

Background events come 
from continuum, $b \rightarrow c \ell \nu$,
$b \rightarrow u \ell \nu$~(other than $B \to \rho \ell \nu$
or $B \to \omega \ell \nu$),
and events with hadrons misidentified as leptons.
The $b \rightarrow c \ell \nu$ background is dominant at lower 
lepton energy.  To distinguish signal events from background
most effectively, 
we consider three lepton-energy ranges:
\begin{itemize}
\setlength{\itemsep}{0.01in}
\item $E_{\ell} > 2.3$~GeV (denoted as HILEP in the
remainder of this paper),
\item $2.0 < E_{\ell} < 2.3$~GeV (LOLEP),
\item $1.7 < E_{\ell} < 2.0$~GeV,
\end{itemize}
where the lepton energy, $E_{\ell}$, is computed in the $\Upsilon(4S)
\to B \bar B$ rest frame (lab frame).  The $B$-meson
rest frame is moving with a small velocity, $\beta \approx 0.065$,
relative to the $\Upsilon(4S)$ frame.
Sensitivity to
$B \rightarrow \rho \ell \nu$ decay is best in the 
HILEP lepton-energy bin where the 
$b \rightarrow c \ell \nu$ background is quite small.
Here, the backgrounds are primarily
continuum
and $b \rightarrow u \ell \nu$ events.  In LOLEP, 
the $b \rightarrow c \ell \nu$ background dominates, but
the $b \rightarrow u \ell \nu$ contributions in this region
are not negligible.
Events with $1.7 < E_{\ell} < 2.0$~GeV are completely dominated
by $b \to c \ell \nu$ decays and are included primarily
to verify our understanding of this source of background.
  
Because our best sensitivity to $B \to \rho \ell \nu$ is in
the HILEP region, suppression of continuum background is a
key issue in the analysis.  At the
$\Upsilon(4S)$ center-of-mass energy,
$\Upsilon(4S) \to B \bar B$ decays can be distinguished from
continuum events, which have a three times larger cross section,
on the basis of various quantities that describe the 
overall distribution of tracks and photons in the event.  Such
event shape, or event topology, variables exploit the
fact that the $B$ mesons are produced nearly at
rest so their decay products 
are distributed roughly uniformly in solid angle.  In contrast,
continuum events yield a much more collimated or jet-like 
event topology.  We use several
event-shape requirements in selecting our sample:

\begin{itemize}
\setlength{\itemsep}{0.01in}
\item The ratio of the second to zeroth Fox-Wolfram
moments~\cite{r2gl} is required to be less
than $0.3$.  This ratio of moments tends to be close to zero for
spherical events and up to unity for jet-like events.
\item $\rho \ell$, $\omega \ell$, and $\pi \ell$
candidates are required to have $\vert \cos \theta_{{\rm thrust}}
\vert \leq 0.8$, where $\theta_{{\rm thrust}}$ is defined to
be the angle between the thrust axis of the candidate system and
the thrust axis of the rest of the event.
\item We select events with $\vert \cos \theta_{{\rm miss}} \vert \leq 0.9$,
where $\theta_{{\rm miss}}$ is the polar angle of the
missing momentum (${\bf p}_{{\rm miss}}$) with respect to the beam axis (see 
Eq.~\ref{eq:pmiss}).
\item We select events whose energy is evenly distributed
around the momentum axis of the candidate lepton
using a Fisher discriminant~\cite{fisher}.
The input variables to the Fisher discriminant are the track 
and CsI cluster energies in nine cones of equal solid angle
around the lepton-momentum axis.
Energy in $B \bar B$ events
tends to be more evenly distributed in these
cones than in jet-like continuum
events.
\end{itemize}

\begin{figure}
\begin{center}
\begin{turn}{-90}
\epsfig{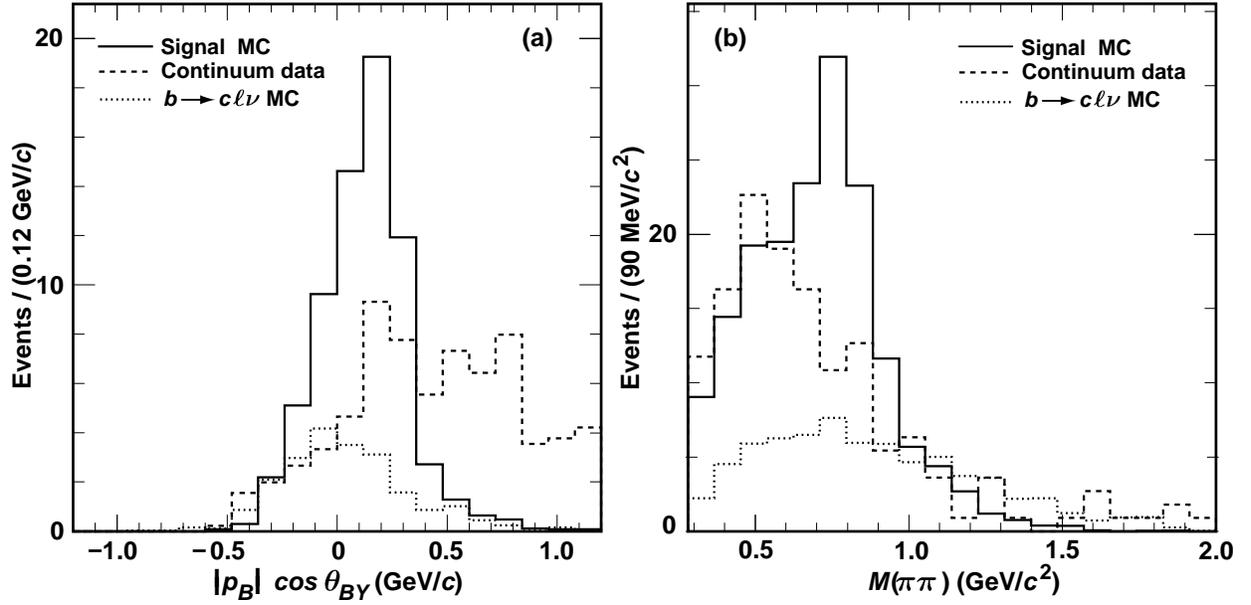}
\end{turn}
\end{center}
\caption{Signal and background distributions for the
HILEP $\pi \pi$ modes: (a) 
$\vert {\bf p}_{B} \vert \cos \theta_{{\rm BY}}$ 
for events that pass all other requirements including
$\vert M(\pi \pi) - M(\rho)\vert  < 0.15~{\rm GeV}/c^2$ and
$\vert \Delta E \vert < 0.5~{\rm GeV}$ and (b)
$M(\pi \pi)$ for events that pass all requirements
including $\vert \Delta E \vert < 0.5~{\rm GeV}$.
Shown are the Monte Carlo signal, continuum data,
and Monte Carlo $b \to c \ell \nu$ contributions.  Each has been
normalized using the results of the likelihood
fit described in Sec.~\protect\ref{sec:anal}. 
The signal contribution includes
events from one signal mode that have been reconstructed 
in another signal mode.  The
$\vert {\bf p}_{B} \vert \cdot \vert \cos \theta_{ BY} \vert  
\leq 385$~MeV/$c$ requirement keeps almost all signal
events, while rejecting much of the background.}
\label{fig:cosby}
\end{figure}

To suppress combinatoric backgrounds from $b \rightarrow c \ell \nu$
and from $b \rightarrow u \ell \nu$ sources other than 
$B \rightarrow \rho \ell \nu$ and $B \to \omega \ell \nu$, 
we require that the
event kinematics be consistent with these signal modes.
Using the constraints
$E_{{B}} = E_{{\rm Beam}}$ and $p_{\nu}^2 = 0 = 
(p_B - p_{\rho} - p_{\ell})^2$, we compute 
the angle between the $B$  momentum
direction and that of the reconstructed
${Y} \equiv (\rho,~\omega,~\pi) + \ell$
system:
\begin{equation}
\cos \theta_{{ BY}} \equiv {{2E_{B} E_{Y}- 
m_{B}^2 - m_{Y}^2   } \over { 2 \vert {\bf p}_{B}
\vert \vert {\bf p}_{Y} \vert}}.
\label{eq:cosby}
\end{equation}
The distribution of $ \vert {\bf p}_{B} \vert \cdot \cos \theta_{ BY}$ 
is shown in Fig.~\ref{fig:cosby}a for
signal, continuum, and $b \to c \ell \nu$ events that satisfy all other
analysis requirements.  We require that 
$\vert {\bf p}_{B} \vert \cdot
\vert \cos \theta_{ BY} \vert 
\leq 385$~MeV/$c$.  For well-reconstructed signal events with
a $B$ momentum of less than 385~MeV/$c$, this cut is $100 \%$
efficient, but $b \rightarrow u \ell \nu$ (non-signal modes),
$b \rightarrow c \ell \nu$, and continuum
events often produce unphysical values of 
$\cos \theta_{{ BY}}$ and can be rejected.  At CESR,
the average $B$ meson momentum is approximately
320~MeV/$c$, with a standard deviation of 
50~MeV/$c$ due to the beam energy spread from the emission of
synchrotron radiation,
so this cut is nearly $100 \%$ efficient for well-reconstructed signal
events.

%%%%%%%%%%%Put figure 1 here

Finally, we compare the direction of the missing momentum 
(${\bf p}_{{\rm miss}}$) with
that of the neutrino momentum (${\bf p}_{\nu}$) 
inferred from the ${\bf p}_{\nu}  = {\bf p}_{B}-{\bf p}_{Y}$ 
momentum.  The latter is known to within an
azimuthal ambiguity about the ${\bf p}_{Y}$
direction because the magnitude, but not the direction, of the
$B$ meson momentum is known.  We
require that the minimum difference in the 
angle between these two directions be
less than 0.6~radians.

Because the $\rho$ has a large width and we are reconstructing
several different
hadronic modes, events with multiple entries passing all of
the above requirements are common.  To avoid the statistical 
difficulties associated with this, we choose one combination
per event after all other criteria are imposed, picking
the combination with $\vert {\bf p}_{Y}+
{\bf p}_{{\rm miss}} \vert $ closest to $\vert {\bf p}_B \vert$.  

The large $\rho$ width also leads to an important effect in
signal events.  The candidate $\pi \pi$ system may not 
consist of the
true daughter particles of the $\rho$.  For example, it is possible for a 
$B \to \rho^0 \ell \nu$ event to satisfy the analysis 
requirements for the $B \to \rho^+ \ell \nu$ channel.
In addition, $B \to \omega \ell \nu$ events can feed into the 
$B \to \rho \ell \nu$ channels.  We denote this contribution
as {\it crossfeed} signal.
Although the kinematic distributions
of these events are somewhat different from correctly
reconstructed signal events, 
the crossfeed contribution is produced by signal processes, 
and it is counted as
such.

Figure~\ref{fig:cosby}b shows the $M(\pi \pi)$ distribution
expected from signal, continuum, and $b \to c \ell \nu$ events
in the HILEP $\pi \pi$ modes.
Each contribution has been normalized using the results of 
the likelihood fit described in Sec.~\ref{sec:anal}. 
  
\section{Fit method}
\label{sec:anal}

We perform
a binned maximum-likelihood fit using MINUIT~\cite{minuit} 
in three lepton-energy ranges and
five decay modes.  In addition to lepton energy, we fit two kinematic
variables, so we are in effect performing a three dimensional correlated fit.
Our fit includes contributions from $B \rightarrow \rho \ell \nu$,
$B \to \omega \ell \nu$,
$B \rightarrow \pi \ell \nu$, other $b \rightarrow u \ell \nu$ modes,
$b \rightarrow c \ell \nu$, continuum events,
as well as a contribution from
hadrons faking leptons.  In this section, we describe the fit variables
and constraints, the modeling of signal and background components, 
and our method for extracting the rate as a function of $q^2$.

The three fit variables, $E_{\ell}$, $M_{\rm had}$, and
$\Delta E$, are constructed from the three decay products in
semileptonic decay: the lepton, the final-state hadron, and the
neutrino. The three lepton-energy bins were discussed in 
Sec.~\ref{sec:events}.
The invariant mass of the daughter hadron, $M_{\rm had}$, is 
reconstructed in the $\pi^+\pi^-$ and $\pi^{\pm}\pi^0$ channels
for $B\to\rho\ell\nu$ and in the $\pi^+\pi^-\pi^0$  channel
for $B\to\omega\ell\nu$.  The bin size for
the fit is $90$~MeV/$c^2$ for both $M(\pi \pi)$ and 
$M(\pi \pi \pi)$.
The momentum of
the hadron is not used as 
a fit variable but is the basis for measuring the
$q^2$ distribution after the fit is performed.

The presence of a neutrino consistent with
$B \to \rho \ell \nu$ decay is signaled by
a peak in the distribution of
\begin{equation}
\Delta E \equiv E_{\rho} + E_{\ell} + \vert {\bf p}_{{\rm miss}} \vert
- E_{{\rm Beam}}
\end{equation}
at $\Delta E = 0$.
Backgrounds from $b \rightarrow c \ell \nu$ and $b \rightarrow u \ell \nu$
(except for $B \to \pi \ell \nu$) events tend to have $\Delta E < 0$.  
On the other hand, 
$B \rightarrow \pi \ell \nu$ events have $\Delta E > 0$ when reconstructed
in the $\pi \pi$ or $\pi^+ \pi^- \pi^0$ 
modes, since extra particles beyond the actual
$B$ decay products are included in the decay hypothesis. 
The $\Delta E$ bin size for the fit is 200~MeV.

\begin{figure}
\begin{center}
\epsfig{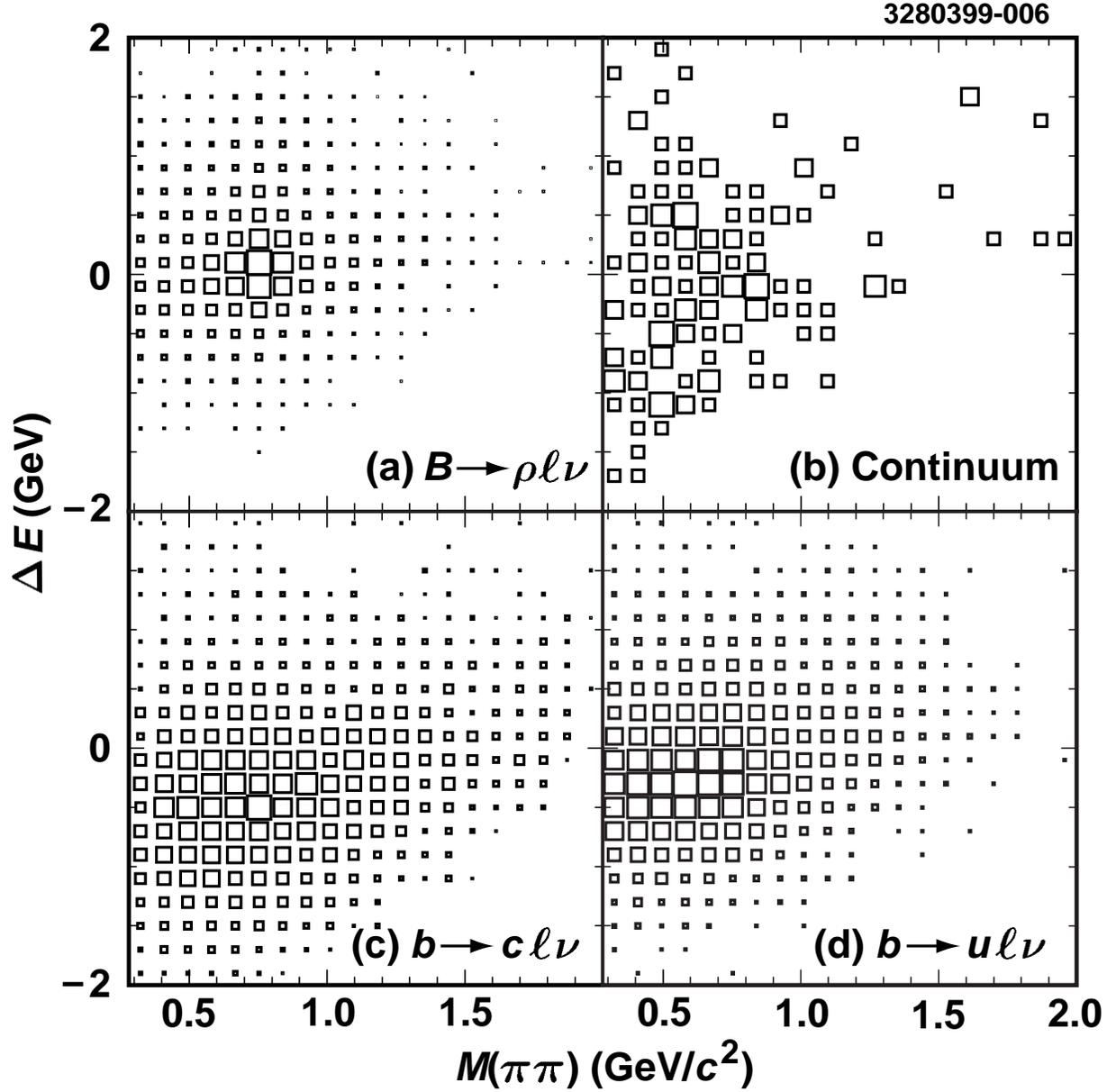}
\end{center}
\caption{Signal and background 
$\Delta E$ vs.~$M(\pi \pi)$  distributions for
the combined $\pi^{\pm} \pi^0$ and $\pi^+ \pi^-$ modes in HILEP:
(a) the direct signal contribution;  
background events from
(b) the continuum; (c) $b \to c \ell \nu$ events;
(d) $b \to u \ell \nu$ events.  The continuum
contribution is modeled using the off-resonance data, while
the other contributions are taken from Monte Carlo samples.
Signal events are
centered on $\Delta E=0$ and are clustered around
the $\rho$ mass in $M(\pi \pi)$.} 
\label{fig:2d}
\end{figure}

$\Delta E$ vs.~$M(\pi \pi)$ distributions for signal and
background contributions in the HILEP region
are shown in Fig.~\ref{fig:2d}.
Shapes of distributions for signal, 
$b \to c \ell \nu$,
and $b \to u \ell \nu$ are taken from 
Monte Carlo simulations using theoretical models of
$B$-meson decays and full detector
simulations based on GEANT~\cite{geant}, while the
continuum background is measured with the off-resonance 
data.  An additional background, not shown, arises from
hadrons passing the lepton identification requirements
in $B \bar B$ events.  This contribution is determined
using measured fake rates and data, as described below.
The data or Monte Carlo samples that are
used to estimate the non-continuum 
contributions are sufficiently large
that their statistical fluctuations are negligible compared
with those of the data.
We now discuss each of these fit components in more detail.

%%%%%%%%%%Put figure 4 here

For signal, the Monte Carlo sample provides
the {\it shapes} of the three-dimensional distribution of 
$E_{\ell}$, $M_{{\rm had}}$, and $\Delta E$.
There is very little model dependence in the $\Delta E$ vs.~$M_{{\rm had}}$
distributions, but the distribution of signal events
across the three lepton-energy bins varies significantly 
among form-factor models.  The 
variation among the $B \rightarrow \rho \ell \nu$ form-factor models
is a significant systematic error for our measurement of
${\cal B}(B \rightarrow \rho \ell \nu)$ and $\vert V_{ub} \vert$.  

There are a large number of crossfeed
signal events, as described in the previous section.
The size of the
crossfeed contribution relative to the direct signal is 
determined using Monte Carlo simulation.
These events are misreconstructed signal decays and
their kinematic distributions 
can be somewhat different from direct signal events. 

Isospin and quark-model relations are used to constrain the relative 
normalizations of $B^0 \rightarrow \rho^- \ell^+ \nu$, 
$B^+ \rightarrow \rho^0 \ell^+ \nu$, and $B^+ \rightarrow \omega
\ell^+ \nu$ and, separately, those of $B^0 \rightarrow \pi^- \ell^+ \nu$ and
$B^+ \rightarrow \pi^0 \ell^+ \nu$.  Isospin symmetry implies that
\begin{eqnarray}
\Gamma(B^0 \rightarrow \rho^- \ell^+ \nu) &=& 2
\Gamma(B^+ \rightarrow \rho^0 \ell^+ \nu), \nonumber \\
\Gamma(B^0 \rightarrow \pi^- \ell^+ \nu) &=& 2
\Gamma(B^+ \rightarrow \pi^0 \ell^+ \nu),
\label{eq:iso}
\end{eqnarray}
while the $\omega$ and $\rho^0$ wave functions are expected 
to be very similar in the quark model, giving
\begin{equation}
\Gamma(B^+ \rightarrow \rho^0 \ell^+ \nu) = 
\Gamma(B^+ \rightarrow \omega \ell^+ \nu). 
\end{equation}
Possible isospin breaking effects are discussed 
in Refs.~\cite{romix,mythesis}.
We assume that the $\Upsilon(4S)$ decays only to
$B \bar B$ in equal proportions of $B^0 \bar B^0$
and $B^+ B^-$ mesons ($f_{+-}/f_{00} = 1.0 \pm 0.1$).  We also
use the value
$\tau_{B^+}/ \tau_{B^0} = 1.04 \pm 0.04$~\cite{pdg} for the 
ratio of charged to neutral $B$-meson lifetimes.

The shape and normalization of the
continuum background are measured using
the off-resonance data sample.
The statistical fluctuations in the off-resonance sample are accounted 
for by fitting the on-
and off-resonance samples simultaneously, accounting for 
the difference in luminosity and cross section.  For each 
bin in the fit, the mean number of continuum background 
events is determined so as to maximize the combined likelihood.

The normalization of the $b \to c \ell \nu$
background is determined to a large extent from the data.
This background arises primarily
from two extensively studied decays, $B \to D \ell \nu$
and $B \to D^* \ell \nu$, and the associated kinematic
distributions are well known. As for the
signal modes, the shapes of the 
$\Delta E$ vs.~$M_{{\rm had}}$ distributions 
due to $b \to c \ell \nu$ are
obtained using the Monte Carlo calculations. With these
constraints, there remain 15 $b \to c \ell \nu$ normalization
parameters, one for each of the five signal
modes in each of three lepton-energy bins.

In our fit, there are in fact only seven
free parameters associated with the $b \to c \ell \nu$
background, defined in the following way.
Five of the parameters, one for each signal
mode, are scale factors that give the overall normalization of the
$b \to c \ell \nu$ background relative to that expected
from the Monte Carlo simulation. Two additional 
parameters describe the ratios of scale factors among 
the three lepton-energy bins. A common set of 
ratios is used for all five modes, so we have
two rather than $2\times5$ parameters for the
$b \to c \ell \nu$ lepton energy spectrum. The reason
for this choice is that the same set of
background modes, dominated by
$B \to D^* \ell \nu$ and $B \to D \ell \nu$, 
contributes to each of the
five signal modes.  The relative contributions of the different
$b \to c \ell \nu$ modes are essentially independent
of the $b \to u \ell \nu$ mode we reconstruct.
In Sec.~\ref{sec:fitresults},
we show that our fit results for all seven
of these $b \to c \ell \nu$ background parameters
agree well with Monte Carlo predictions.

The normalization for the 
contribution from $B \to \pi \ell \nu$
decay is also determined by the data.  As for the $B \to \rho \ell \nu$
contribution, form-factor models are used to describe
the kinematic distributions.  A systematic
error is assigned based on the variation observed 
in the $B \to \rho \ell \nu$ fit results for different
$B \to \pi \ell \nu$ form-factor models.

The ISGW2~\cite{isgw2} predictions are used to model 
the distributions of $b \rightarrow u
\ell \nu$ sources other than
$B \to \rho \ell \nu$,
$B \to \omega \ell \nu$, and
$B \to \pi \ell \nu$
for resonances up to the $\rho(1450)$.
The dominant resonances are $\eta$, $b_1$, and $a_1$.
The uncertainty on the composition of this background
is considered as a systematic error.  To reduce this
systematic uncertainty, we allow for a separate 
normalization for the $b \to u \ell \nu$ component in
each lepton-energy bin.

The small contribution
from fake leptons is found using the data.  The shapes
of the $\Delta E$ vs.~$M(\pi \pi)$ distribution, along
with that of $E_{\ell}$, are found by combining $\rho$ 
candidates with charged tracks not identified as leptons.  These 
combinations are required to pass all analysis
criteria except for the lepton-identification requirements.  
The normalization of the fake-lepton component is determined
by measurements of the probability for hadrons to satisfy 
the lepton identification requirements using $\pi$, $K$,
and $p$ samples from the data (see Sec.~\ref{sec:events}).

There are thus twelve free parameters in the fit:
\begin{itemize}
\setlength{\itemsep}{0.01in}
\item ${\cal B}(B \rightarrow \rho \ell \nu)$ (1 parameter).
\item ${\cal B}(B \rightarrow \pi \ell \nu)$ (1 parameter).
\item The yield of $b \rightarrow u \ell \nu$ events in
each lepton-energy bin (3 parameters).
\item The seven parameters that describe the normalization of the
$b \rightarrow c \ell \nu$ contribution.
\end{itemize}

We do not include a contribution from
$B \to \pi \pi \ell \nu$ nonresonant events
in our fit.  Instead, we use the data to
constrain the size of this contribution and
assign a systematic error (see Sec.~\ref{sec:sys}).

The ${\cal B}(B \rightarrow \rho \ell \nu)$ result from
the likelihood fit is used to compute 
\begin{equation}
\vert V_{ub} \vert = 
\sqrt{ {   {\cal B} \over  
{\tilde{\Gamma}_{{\rm thy}} \tau_{B^0}}}}.
\label{eq:vub}
\end{equation}
${\cal B}$ is the result for the
$B^0 \to \rho^- \ell^+ \nu$ branching fraction from the likelihood
fit, which depends on the form-factor model used to
describe the signal distributions.  Results for
$\tilde{\Gamma}_{{\rm thy}}$ are given for each of the five
form-factor models in Table~\ref{table:q2info}.  
We use $\tau_{B^0} = 1.56 \pm 0.04$~ps~\cite{pdg}.

The likelihood fit results are also used to 
measure the $q^2$ distribution in 
$B \to \rho \ell \nu$ decay.  The large 
$b \to c \ell \nu$ backgrounds in LOLEP limit
this measurement to the HILEP lepton-energy region.  
To determine the $q^2$ distribution,
we consider events in
the $\pi^{\pm} \pi^0$ and $\pi^+ \pi^-$ modes with \
$\vert M(\pi \pi) - M(\rho) \vert < 0.15$~GeV/$c^2$ and
$\vert \Delta E \vert < 0.5$~GeV.
The background contributions in each $q^2$ interval
are estimated using the results of the likelihood
fit.  We count the number of
events in the background samples and 
subtract them from the data.
Below we discuss the determination
of $q^2$ for each event and describe the extraction
of the $q^2$ distribution.

\begin{figure}
\begin{center}
\epsfig{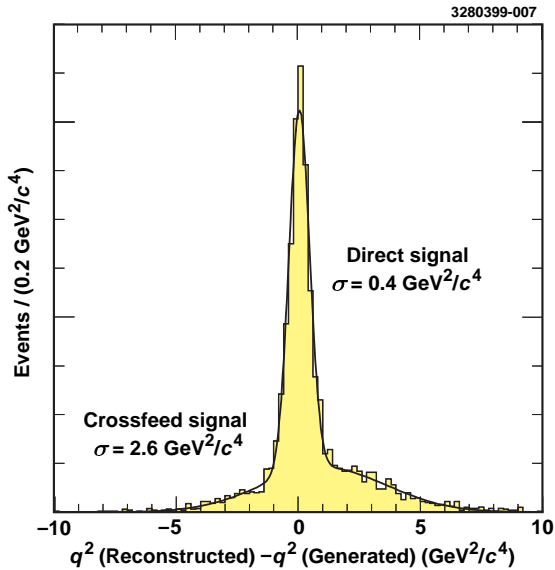}
\end{center}
\caption{The $q^2$ resolution for signal Monte Carlo events.  
The curve shows a fit to the Monte Carlo events (shaded
histogram) using two Gaussians.
The narrow ($\sigma = 0.4$~GeV$^2/c^4$)
component is due to events where the correct hadronic mode is 
reconstructed.  The broader ($\sigma = 2.6$~GeV$^2/c^4$) component
is from crossfeed events.
We account for this smearing in our measurement of $\Delta \Gamma$ in bins
of $q^2$.  The full $q^2$ range for $B \to \rho \ell \nu$ is
$\Delta q^2 \approx 21$~GeV$^2/c^4$.}
\label{fig:q2resol}
\end{figure}

If the $B$ meson momentum vector were known, $q^2$
could be determined unambiguously from the energy of
the daughter hadron.  For $B \to \rho \ell \nu$,
\begin{equation}
q^2=M_B^2+M_{\rho}^2- 2E_B E_{\rho} + 
2 {\bf p}_B \cdot {\bf p}_{\rho}.
\end{equation}
We know the 
magnitude of the $B$ momentum, but not its direction,
which introduces an uncertainty in $q^2$.  Using
the beam-energy constraint, we compute the minimum
and maximum allowed $q^2$ values for each $\rho+\ell$ 
combination.  These correspond to the kinematic 
configurations where the $B$ momentum direction is
closest to, or furthest from, the $\rho$ momentum direction.
The midpoint of the allowed $q^2$ 
range is our estimate of $q^2$. The $q^2$ resolution function
is shown in Fig.~\ref{fig:q2resol}.  

To find the true $q^2$ distribution from the 
data, we must determine the efficiency of our 
analysis requirements as a function of $q^2$, including the
smearing effects shown in Fig.~\ref{fig:q2resol}.
We perform this measurement in three equal bins
of $q^2$.  
Monte Carlo simulation
is used to compute the efficiency of events generated
with a particular $q^2$ to be reconstructed at a different
value of $q^2$.
This efficiency matrix is computed for
each of the five form-factor models that we consider; model
dependence arises primarily from the 
efficiency of the
$E_{\ell} > 2.3$~GeV requirement.  
The reconstructed $q^2$ distribution of the
HILEP data and this efficiency
matrix are used to determine the true $q^2$
distribution.  

%%%%%%%%%%Put figure 2 here

\section{Fit results}
\label{sec:fitresults}

The signal and background yields extracted from
the fit in HILEP and LOLEP 
are shown in Tables~\ref{table:yields} and~\ref{table:sigyields} for the $\pi \pi$ and
$\pi^+ \pi^- \pi^0$ modes.
The signal yields for the combined $\pi^+ \pi^0$
and $\pi^+ \pi^-$ modes in HILEP, where our sensitivity to 
the signal is best, are $123 \pm 18$ (direct) and 
$93 \pm 14$ (crossfeed),
for a total of $216 \pm 32$ signal events.  
The five $b \rightarrow c \ell \nu$ normalization scale factors are
consistent with Monte Carlo predictions
to better than $5 \%$.  The two scale factor ratios that determine the 
lepton-energy distribution 
for the $b \to c \ell \nu$ contribution are within one standard deviation
of unity.  The 
$b \rightarrow u \ell \nu$ (modes other than
$B \to \rho \ell \nu$, $B \to \omega \ell \nu$, and $B \to \pi \ell \nu$)
normalizations in the three lepton-energy bins
agree with each other to better than $20 \%$, 
within one standard deviation,
indicating
that the ISGW2 cocktail of $b \rightarrow u \ell \nu$ 
modes is adequate, at least to
within our sensitivity.  

\begin{table}
\begin{center}
\begin{tabular}{ccccccc}
                                         & $\rho^{\pm}$ HI &$\rho^0$ HI   &$\omega$ HI   & $\rho^{\pm}$ LO &$\rho^0$ LO    &$\omega$ LO    \\ \hline
$\Upsilon({\rm 4S})$ yield               & 198             & 621          & 460          & 2249            & 7298          & 8552          \\
$e^+ e^- \rightarrow q \bar q$ bkg.                           &$63 \pm 11$      & $248 \pm 22$ & $250 \pm 22$ & $127 \pm 16$    & $398 \pm 28$  & $437 \pm 29$  \\
$b \rightarrow c \ell \nu$ bkg.          &$39 \pm 5$       & $90 \pm 7$  & $52 \pm 5$   & $1941 \pm 47$   & $6419 \pm 114$ & $7804 \pm 129$ \\
fake lepton bkg.                         &$4 \pm 2$        & $13 \pm 7$   & $10 \pm 5$   & $11 \pm 7$      & $35 \pm 21$   & $44 \pm 22$   \\
$b \rightarrow u \ell \nu$ bkg. &$25 \pm 10$       & $57 \pm 23$  & $43 \pm 17$  & $97 \pm 24$    & $222 \pm 57$  & $236 \pm 60$  \\
$B \rightarrow \pi \ell \nu$  bkg.       &$9 \pm 3$       & $20 \pm 6$   & $4 \pm 1$    & $15 \pm 5$      & $32 \pm 10$    & $6 \pm 2$     \\ \hline
Direct sig.                   &$56 \pm 8$       & $67 \pm 10$   & $21 \pm 3$   & $61 \pm 9$      & $79 \pm 12$    & $31 \pm 5$    \\
Xfeed sig.                       &$13 \pm 2$       & $80 \pm 12$  & $32 \pm 5$   & $23 \pm 3$      & $100 \pm 15$   & $52 \pm 8$    \\ 
\end{tabular}
\end{center}
\caption{Summary of data yields for the $\rho^{\pm}$, $\rho^0$, and $\omega$
modes with lepton energies between 2.3 and 2.7~GeV (HILEP) and
between 2.0 and 2.3~GeV (LOLEP). 
Yields of the background 
contributions are insensitive
to the form-factor model used to model $B \rightarrow \rho \ell \nu$
events; the yields presented in this table were
obtained using the UKQCD model.  The $b \rightarrow u \ell \nu$
background includes all $B \rightarrow X_u \ell \nu$ modes except for
$\rho$, $\omega$, and $\pi$.
The crossfeed (Xfeed) signal contribution corresponds to
events from one vector mode passing the selection cuts of another
vector mode ($\rho^0 \ell \nu \leftrightarrow \rho^{\pm} \ell \nu$
or $\rho \ell \nu \leftrightarrow \omega \ell \nu$)
and is constrained to the signal in the fit.  
All errors are statistical only.
The errors on the
direct and crossfeed signal yields  
are completely correlated, as are those on the yields of 
$B \rightarrow \pi \ell \nu$ events. 
Errors on the $b \rightarrow u \ell \nu$
background are completely correlated among modes, but not between HILEP 
and LOLEP. }
\label{table:yields}
\end{table}

\begin{table}
\begin{center}
\begin{tabular}{ccccccc}
   & $\rho^{\pm}$ HI &$\rho^0$ HI   &$\omega$ HI   & $\rho^{\pm}$ LO &$\rho^0$ LO    &$\omega$ LO    \\ \hline
Signal eff. (ISGW2)      &  0.040    & 0.091     & 0.032     & 0.039     & 0.094     & 0.040   \\ 
Signal yield   (ISGW2)                   &$58 \pm 9$       & $68 \pm 10$   & $21 \pm 3$   & $55 \pm 9$      & $70 \pm 11$    & $26 \pm 4$    \\
Xfeed yield (ISGW2)                      &$14 \pm 2$       & $80 \pm 12$  & $34 \pm 5$   & $20 \pm 3$      & $89 \pm 13$   & $48 \pm 7$    \\ \hline
Signal eff. (LCSR)       & 0.026     & 0.061     & 0.022     & 0.030     & 0.076     & 0.034    \\ 
Signal yield  (LCSR)                     &$56 \pm 8$       & $66 \pm 10$   & $21 \pm 3$   & $64 \pm 10$      & $84 \pm 13$    & $33 \pm 5$    \\
Xfeed yield (LCSR)                      &$14 \pm 2$       & $80 \pm 11$  & $32 \pm 5$   & $24 \pm 3$      & $105 \pm 15$   & $54 \pm 8$    \\ \hline
Signal eff. (UKQCD)      & 0.031     & 0.070     & 0.025     & 0.033     & 0.083     & 0.036    \\ 
Signal yield  (UKQCD)                   &$56 \pm 8$       & $67 \pm 10$   & $21 \pm 3$   & $61 \pm 9$      & $79 \pm 12$    & $31 \pm 5$    \\
Xfeed yield (UKQCD)                      &$14 \pm 2$       & $80 \pm 12$  & $33 \pm 5$   & $23 \pm 3$      & $100 \pm 15$   & $52 \pm 8$    \\ \hline
Signal eff. (Wise/Ligeti+E791)& 0.035& 0.080     & 0.028     & 0.037     & 0.094     & 0.040     \\ 
Signal yield  (Wise/Ligeti+E791)         &$56 \pm 8$       & $66 \pm 10$   & $21 \pm 3$   & $60 \pm 9$      & $79 \pm 12$    & $30 \pm 4$    \\
Xfeed yield (Wise/Ligeti+E791) & $14 \pm 2$ & $80 \pm 12$  & $32 \pm 5$ & $24 \pm 3$ & $102 \pm 15$   & $52 \pm 8$    \\ \hline
Signal eff. (Beyer/Melikhov)  & 0.029& 0.067     & 0.024     & 0.033     & 0.083     & 0.036     \\ 
Signal yield (Beyer/Melikhov)            &$55 \pm 8$       & $66 \pm 10$   & $21 \pm 3$   & $61 \pm 10$      & $81 \pm 12$    & $31 \pm 5$    \\
Xfeed yield (Beyer/Melikhov)  &$14 \pm 2$& $80 \pm 12$& $32 \pm 5$& $24 \pm 3$& $104 \pm 15$ & $53 \pm 8$    \\ 
\end{tabular}
\end{center}
\caption{Summary of direct and crossfeed (Xfeed) signal yields 
for the $\rho^{\pm}$, $\rho^0$, and $\omega$
modes with lepton energies between 2.3 and 2.7~GeV (HILEP) and
between 2.0 and 2.3~GeV (LOLEP).  Signal efficiencies are
normalized to the number of events in the full lepton-energy range.  
The crossfeed contribution corresponds to
events from one vector mode passing the selection cuts of another
vector mode ($\rho^0 \ell \nu \leftrightarrow \rho^{\pm} \ell \nu$
or $\rho \ell \nu \leftrightarrow \omega \ell \nu$)
and is constrained to the signal in the fit.  
All errors are statistical only.  Errors on the
direct and crossfeed signal yields are completely correlated.}
\label{table:sigyields}
\end{table}

\begin{figure}
\begin{center}
\epsfig{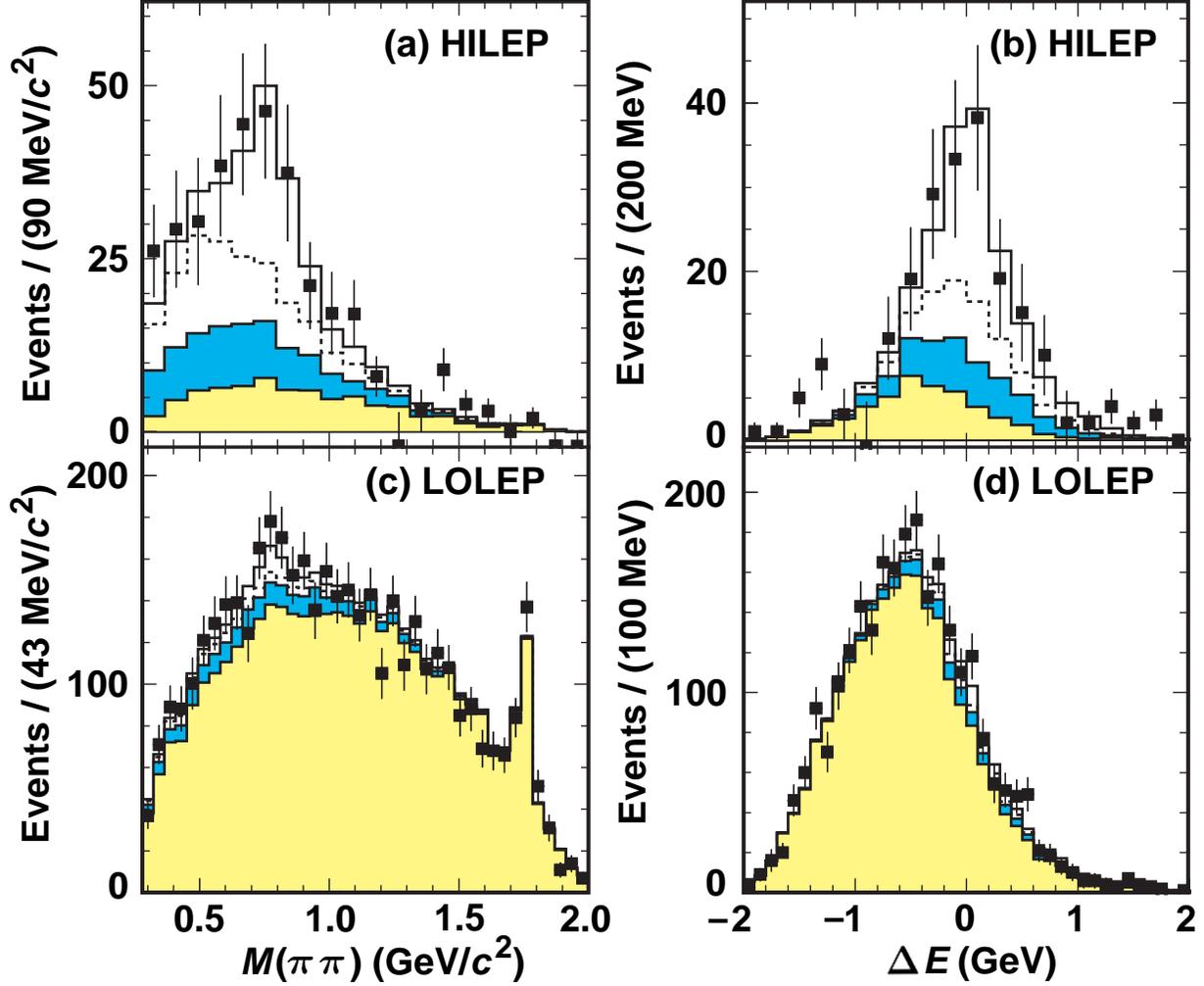}
\end{center}
\caption{
Projections of the fit for the combined $\rho^{\pm}$ and $\rho^0$ modes
for HILEP ($E_{\ell} > 2.3$~GeV, upper plots) and LOLEP 
($2.0 < E_{\ell} < 2.3$~GeV, lower plots):
(a) $M(\pi \pi)$ for HILEP 
after a $\vert \Delta E \vert < 0.5$~GeV
cut; (b) $\Delta E$ for HILEP
after a $\vert M(\pi \pi) - M({\rho}) \vert < 
0.15$~GeV/$c^2$ cut;
(c) $M(\pi \pi)$ for LOLEP 
after a $\vert \Delta E \vert < 0.5$~GeV
cut;
(d) $\Delta E$ for LOLEP
after a $\vert M(\pi \pi) - M({\rho}) \vert < 
0.15$~GeV/$c^2$ cut.
In each plot, the points with error bars show the on-resonance
data after continuum-background subtraction, 
while the histogram shows the
projection of the fit. The contributions to the fit are the direct and
crossfeed components of the signal (unshaded regions, above
and below the dashed line, respectively); 
the background from $b\to u\ell \nu$
non-signal modes (darkly-shaded region); and the background
from $b\to c\ell \nu$ (lightly-shaded region). The $b\to u\ell \nu$
background includes $B\to\pi\ell \nu$ contributions.}
\label{fig:hilepsig}
\end{figure}

\begin{figure}
\begin{center}
\epsfig{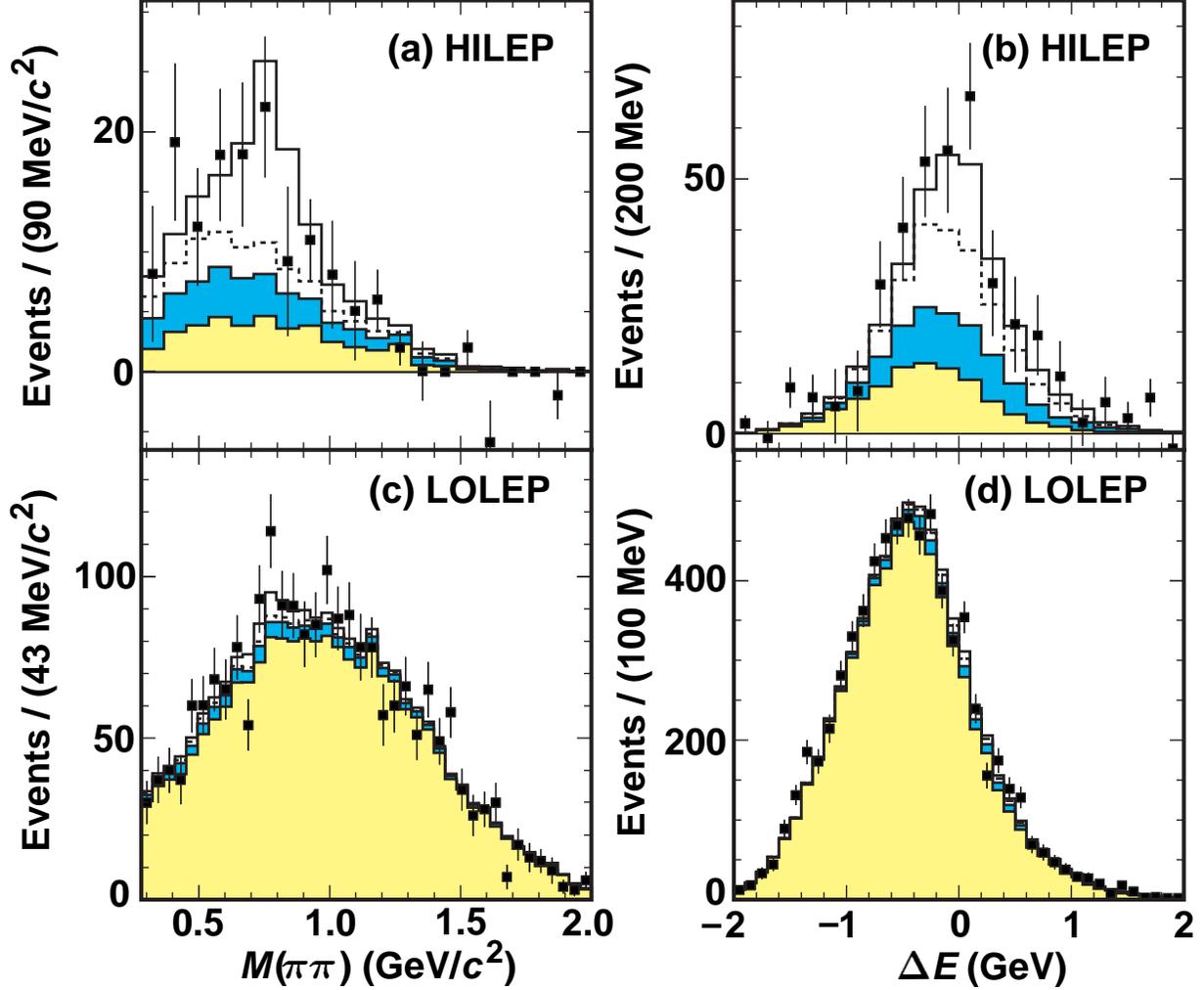}
\end{center}
\caption{
Projections of the fit for the combined $\rho^{\pm}$ and $\rho^0$ modes
for HILEP ($E_{\ell} > 2.3$~GeV, upper plots) and LOLEP 
($2.0 < E_{\ell} < 2.3$~GeV, lower plots):
(a) $M(\pi \pi)$ for HILEP 
after a $\vert \Delta E \vert > 0.5$~GeV cut; 
(b) $\Delta E$ for HILEP after a $\vert M(\pi \pi) - M({\rho}) \vert > 
0.15$~GeV/$c^2$ cut; (c) $M(\pi \pi)$ for LOLEP 
after a $\vert \Delta E \vert > 0.5$~GeV cut;
(d) $\Delta E$ for LOLEP
after a $\vert M(\pi \pi) - M({\rho}) \vert > 
0.15$~GeV/$c^2$ cut.
In each plot, the points with error bars show the on-resonance
data after continuum-background subtraction, 
while the histogram shows the
projection of the fit. The contributions to the fit are the direct and
crossfeed components of the signal (unshaded regions, above
and below the dashed line, respectively); 
the background from $b\to u\ell \nu$
non-signal modes (darkly-shaded region); and the background
from $b\to c\ell \nu$ (lightly-shaded region). 
The $b\to u\ell \nu$
background includes  $B\to\pi\ell \nu$ contributions.}
\label{fig:sidebands}
\end{figure}

We show projections of the fit in both signal- and background-dominated
regions of the $\Delta E$ vs.~$M(\pi \pi)$ distributions.  
Figure~\ref{fig:hilepsig} shows HILEP and LOLEP projections 
onto $M(\pi \pi)$ for $\vert \Delta E \vert < 0.5$~GeV and 
onto $\Delta E$ for
$\vert M(\pi \pi) - M(\rho) \vert < 0.15$~GeV/$c^2$.
We observe a significant $B \to \rho \ell \nu$ signal
in the $M(\pi \pi)$ and $\Delta E$ distributions in HILEP.
Figure~\ref{fig:sidebands} shows the same distributions for
sidebands with $\vert \Delta E \vert > 0.5$~GeV and 
$\vert M(\pi \pi) - M(\rho) \vert > 0.15$~GeV/$c^2$, where we
expect much less signal.  
In both signal- and background-dominated distributions, we
observe good agreement between the data and
the fit projections.
In the projections shown, the signal component is modeled
using the ISGW2 form-factor model, and the continuum
contribution has been subtracted.  

%%%%%%%%%%Put figure 4 here

%%%%%%%%%%Put figure 5 here

The $b \rightarrow c \ell \nu$
background is quite small in the HILEP plots but is dominant in
all of the LOLEP projections.  The peak in the LOLEP $\pi \pi$
modes at large $M(\pi \pi)$ is due to $B \rightarrow D \ell \nu$,
with $D \rightarrow K \pi$, where the $K$ has been misinterpreted
as a pion.  LOLEP projections in Fig.~\ref{fig:sidebands} show
that the shape of the $b \to c \ell \nu$ Monte Carlo distribution
describes the data well in regions of the fit where the
$b \to u \ell \nu$ contributions are small.
In Fig.~\ref{fig:hilepsig}, 
the shape of the $M(\pi \pi)$ distribution in 
the LOLEP lepton-energy bin
indicates that the $b \rightarrow u \ell \nu$ contributions
are necessary to describe properly the data below the 
$b \rightarrow c \ell \nu$ lepton-energy endpoint region.
We do not show fit projections for the data with $E_{\ell} < 2.0$~GeV,
but the fit agrees well with the data in this region.
The $b \to c \ell \nu$ Monte Carlo simulates $M(\pi \pi)$ 
and $\Delta E$ projections of the data well in both shape
and normalization.  

\begin{figure}
\begin{center}
\epsfig{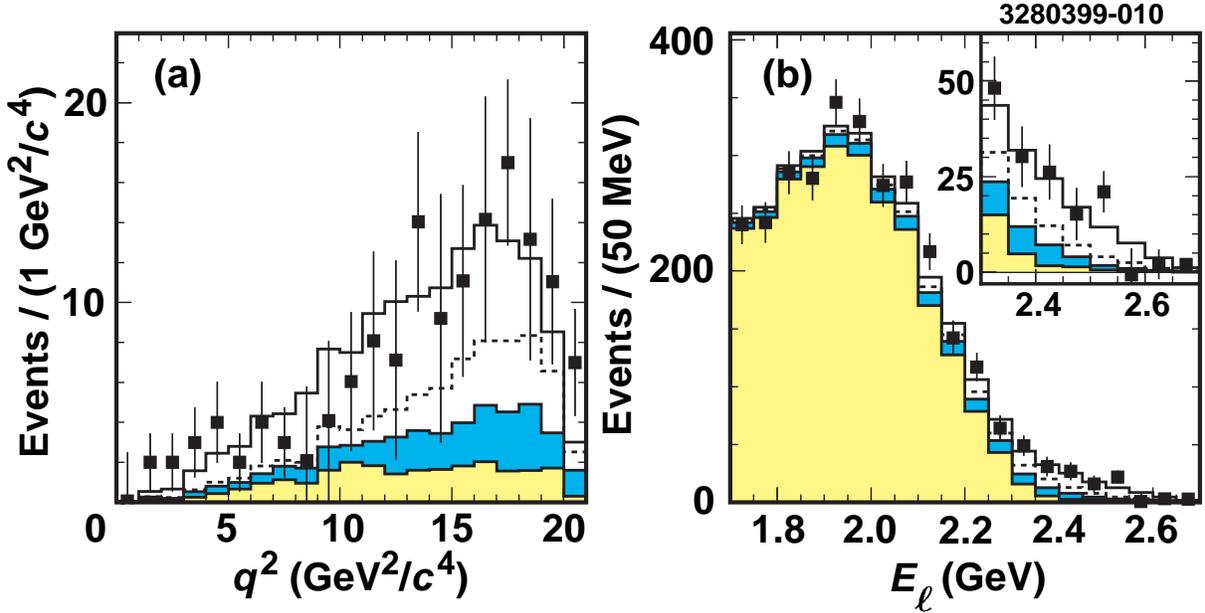}
\end{center}
\caption{Projections of the kinematic variables: (a) $q^2$ and
(b) $E_{\ell}$.
Both distributions are shown for the combined $\rho^{\pm}$ and $\rho^0$
modes for events with $\vert \Delta E \vert < 0.5$~GeV and 
$\vert M(\pi \pi) - M({\rho}) \vert < 0.15$~GeV/$c^2$.
The $q^2$ distribution is shown only for events
with $E_{\ell} > 2.3$~GeV. 
In each plot, the points with error bars show the on-resonance
data after continuum-background subtraction, 
while the histogram shows the
projection of the fit. The contributions to the fit are the direct and
crossfeed components of the signal (unshaded regions, above
and below the dashed line, respectively); 
the background from $b\to u\ell \nu$
non-signal modes (darkly-shaded region); and the background
from $b\to c\ell \nu$ (lightly-shaded region). 
The $b\to u\ell \nu$
background includes  $B\to\pi\ell \nu$ contributions.}
\label{fig:hilepmisc}
\end{figure}

Figure~\ref{fig:hilepmisc} shows the 
reconstructed $q^2$ and lepton-energy distributions for the
$\pi \pi$ modes with
$\vert \Delta E \vert < 0.5$~GeV and 
$\vert M(\pi \pi) - M(\rho) \vert < 0.15$~GeV/$c^2$ requirements.  
We observe good agreement between the data and
the $q^2$ distribution predicted in HILEP by
the form-factor models.  The lepton-energy spectrum predicted
from the fit results 
shows good agreement with the data over the full range of
lepton-energy used ($E_{\ell} > 1.7$~GeV).  
A large excess in the lepton-energy spectrum, 
consistent with the fit projection,
is observed above the continuum contribution
in the region beyond the $b \to c \ell \nu$ lepton energy
endpoint.  

%%%%%%%%%%Put figure 6 here

\begin{figure}
\begin{center}
\epsfig{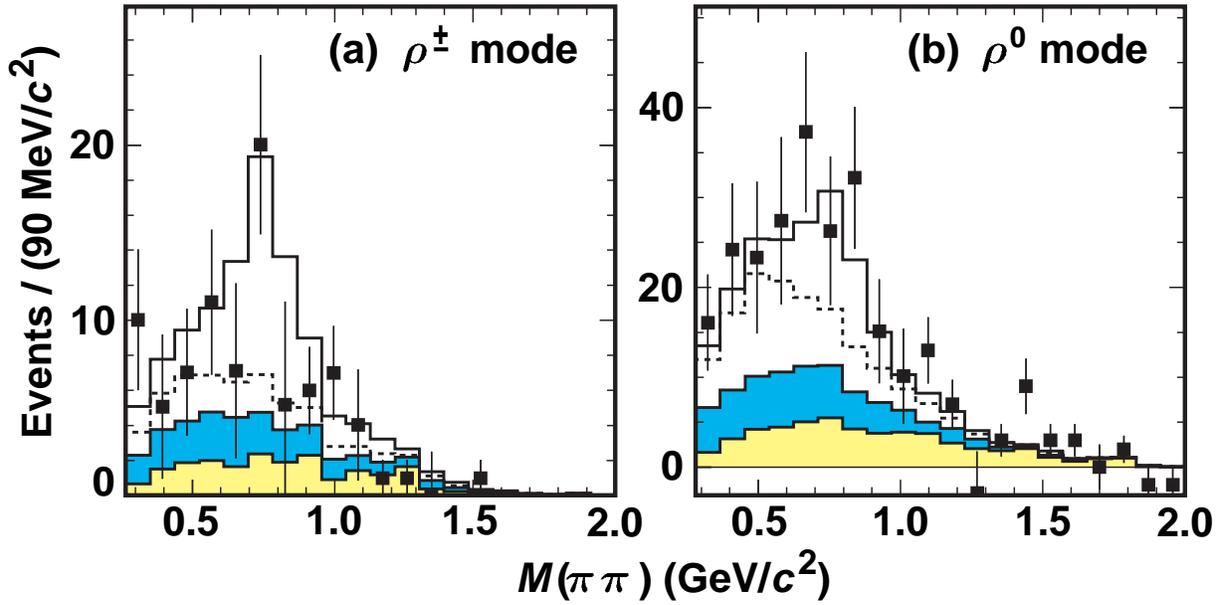}
\end{center}
\caption{
Projections of the fit onto $M(\pi \pi)$
for HILEP 
($E_{\ell} > 2.3$~GeV) with a 
$\vert \Delta E \vert < 0.5$~GeV requirement: (a)
for the $\pi^{\pm} \pi^0$ mode and (b) for the
$\pi^+ \pi^-$ mode.
In each plot, the points with error bars show the on-resonance
data after continuum-background subtraction, 
while the histogram shows the
projection of the fit. The contributions to the fit are the direct and
crossfeed components of the signal (unshaded regions, above
and below the dashed line, respectively); 
the background from $b\to u\ell \nu$
non-signal modes (darkly-shaded region); and the background
from $b\to c\ell \nu$ (lightly-shaded region). The $b\to u\ell \nu$
background includes  $B\to\pi\ell \nu$ contributions.}
\label{fig:mpponemode}
\end{figure}

In the above projections,
the $\pi^{\pm} \pi^0$ and $\pi^+ \pi^-$ modes have been
combined.  Figure~\ref{fig:mpponemode} shows the
$M(\pi \pi)$ distribution with a $\vert \Delta E \vert < 0.5$~GeV
requirement for these modes individually where the
ratio of  
$\Gamma (B^0 \to \rho^- \ell^+ \nu)$
to $\Gamma(B^+ \to \rho^0 \ell^+ \nu)$ has been allowed
to float in the fit.  We find
$\Gamma (B^0 \to \rho^- \ell^+ \nu) /
 \Gamma(B^+ \to \rho^0 \ell^+ \nu) = 1.7^{+1.0}_{-0.6} $, where the
error is statistical only.  This value is in good agreement with
the isospin relation in Eq.~\ref{eq:iso}, which is used to
determine our final results.

%%%%%%%%%%Put figure 6.5 here

\begin{figure}
\begin{center}
\epsfig{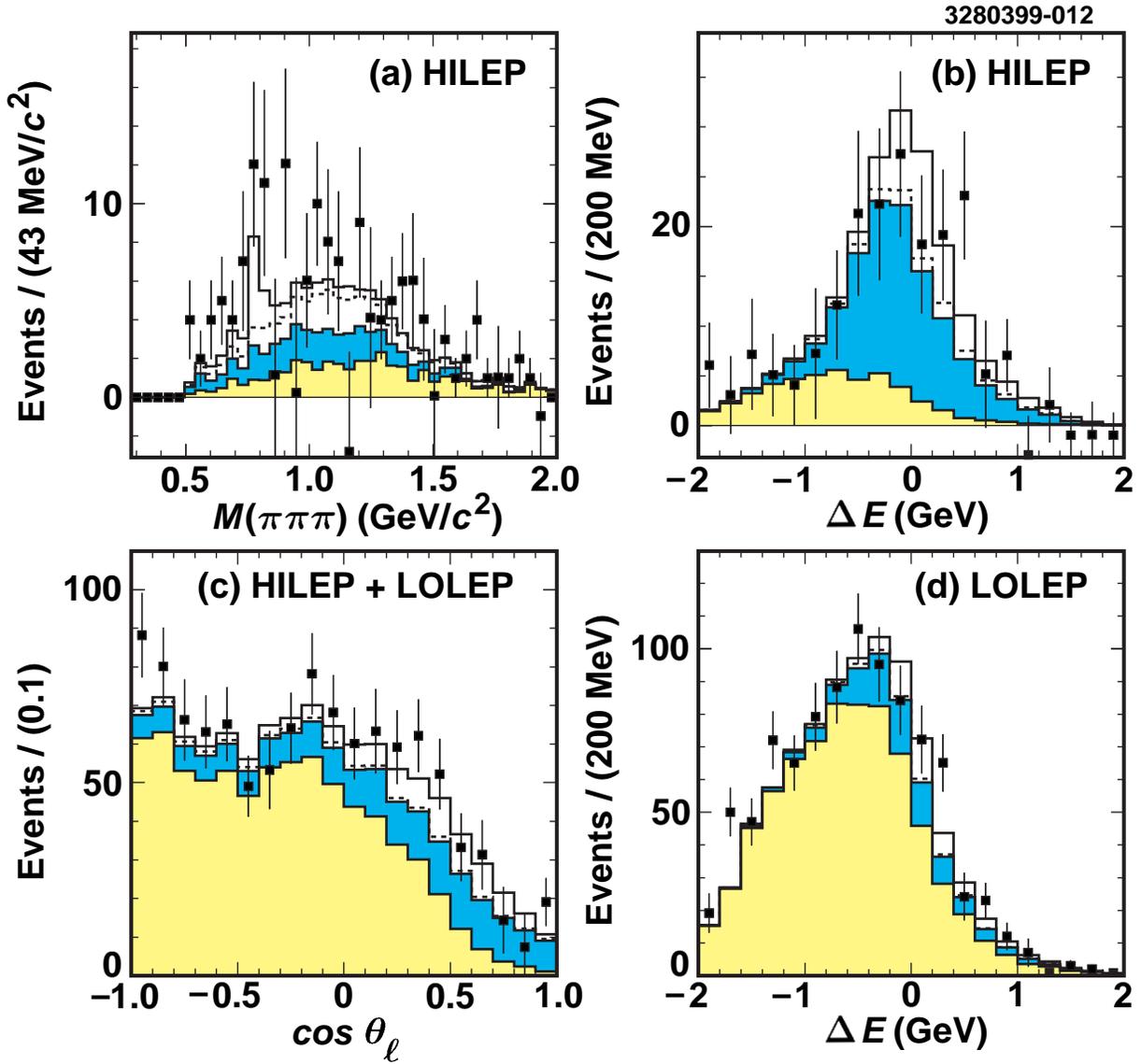}
\end{center}
\caption{Projections of the fit in the $\pi^+ \pi^- \pi^0$
and combined $\pi^{\pm}$ and $\pi^0$ modes in HILEP and LOLEP: (a)
the HILEP $M(\pi^+ \pi^- \pi^0)$
distribution in the $\pi^+ \pi^- \pi^0$ mode with
$\vert \Delta E \vert < 0.5$~GeV;  
(b) the HILEP $\Delta E$ distribution for the $\pi$
modes; (c) the $\cos \theta_{\ell}$ distribution
for the combined HILEP and LOLEP $\pi$ modes;
(d) the LOLEP $\Delta E$ distribution for the $\pi$  
modes.   $\theta_{\ell}$ is
the angle between the lepton momentum direction
in the $W$ rest frame and the $W$ momentum direction
in the $B$ rest frame.
In each plot, the points with error bars show the on-resonance
data after continuum-background subtraction, 
while the histogram shows the
projection of the fit. The contributions to the fit are the direct and
crossfeed components of the signal (unshaded regions, above
and below the dashed line, respectively); 
the background from $b\to u\ell \nu$
non-signal modes (darkly-shaded region); and the background
from $b\to c\ell \nu$ (lightly-shaded region). For the
$B\to\omega\ell \nu$ channels, the $b\to u\ell \nu$
background includes $B\to\pi\ell \nu$ contributions.
For the
$B\to\pi \ell \nu$ channels, the $b\to u\ell \nu$
background includes $B\to\rho (\omega) \ell \nu$ contributions.}
\label{fig:hilepomegapi}
\end{figure}

In Fig.~\ref{fig:hilepomegapi}a
we show the $M(\pi^+ \pi^- \pi^0)$ plot for the HILEP
lepton-energy bin.  We do not observe a significant 
$B \rightarrow \omega \ell \nu$ signal, but the fit
describes the data well.

Figure~\ref{fig:hilepomegapi} also
shows the $\Delta E$ distributions for the HILEP
(\ref{fig:hilepomegapi}b)
and LOLEP (\ref{fig:hilepomegapi}d)
combined $\pi^{\pm}$ 
and $\pi^0$ modes and the $\cos \theta_{\ell}$ 
(\ref{fig:hilepomegapi}c)
distribution for the HILEP and LOLEP $\pi^{\pm}$ and $\pi^0$
modes.  Independent of model, $\theta_{\ell}$ is 
expected to have a $\sin^2 \theta$ distribution
for $B \to \pi \ell \nu$ events.
The HILEP $B \rightarrow \pi \ell \nu$
modes are dominated by $B \rightarrow \rho \ell \nu$ backgrounds.
We find
${\cal B}(B^0 \rightarrow \pi^- \ell^+ \nu) = (1.3 \pm 0.4) \times
10^{-4}$, where the quoted error is statistical only,
consistent with
the previous CLEO result~\cite{lkgprl}.  We do not
quote a full $B \rightarrow \pi \ell \nu$ result as it is very
sensitive to systematics related to
the large $B \rightarrow \rho \ell \nu$ 
backgrounds.

%%%%%%%%%%Put figure 7 here

Table~\ref{table:deltagammafitres} shows the results
of the $\Delta \Gamma$ measurement before the correction
for the lepton-energy cut is performed.  The small spread
seen among form-factor models is due to differences in
the background subtraction and the smearing correction.
Results for the full lepton-energy range are computed
using the predictions in Table~\ref{table:elepeffvsq2} and are 
discussed in Sec.~\ref{sec:res}.

\begin{table}
\begin{center}
\begin{tabular}{cccc}
FF model & $\Delta \Gamma$ ($/10^{-2}~{\rm ns}^{-1}$) 
&$\Delta \Gamma$ ($/10^{-2}~{\rm ns}^{-1}$) 
&$\Delta \Gamma$ ($/10^{-2}~{\rm ns}^{-1}$) \\
 & $0<q^2<7$~GeV$^2/c^4$ &
 $7<q^2<14$~GeV$^2/c^4$ &
 $14<q^2<21$~GeV$^2/c^4$ \\ 
&$E_{\ell}>2.3$~GeV & $E_{\ell}>2.3$~GeV & $E_{\ell}>2.3$~GeV \\ \hline
ISGW2            & 1.2 $\pm$ 0.5 & 1.4 $\pm$ 0.8 & 2.7 $\pm$ 0.8 \\
LCSR             & 1.3 $\pm$ 0.5 & 1.4 $\pm$ 0.8 & 2.7 $\pm$ 0.8\\
UKQCD            & 1.2 $\pm$ 0.5 & 1.4 $\pm$ 0.8 & 2.7 $\pm$ 0.8\\
Wise/Ligeti+E791 & 1.1 $\pm$ 0.5 & 1.4 $\pm$ 0.8 & 2.8 $\pm$ 0.8\\
Beyer/Melikhov   & 1.1 $\pm$ 0.5 & 1.4 $\pm$ 0.8 & 2.8 $\pm$ 0.8\\ 
\end{tabular}
\end{center}
\caption{Results for $\Delta \Gamma(B^0 \to \rho^- \ell^+ \nu)$ in 
bins of $q^2$ for events with $E_{\ell} > 2.3$~GeV 
for each form-factor (FF) model.  The 
errors are statistical only.}
\label{table:deltagammafitres}
\end{table}

\section{Systematic errors}
\label{sec:sys}

Table~\ref{table:sysbr} summarizes our systematic errors.  
The largest systematic errors are due to 
uncertainties on the $b \rightarrow c \ell \nu$
and $b \rightarrow u \ell \nu$ backgrounds, the
dependence of the efficiencies of our selection criteria
on the Monte Carlo modeling of the detector, and the
possible contamination due to $B \to \pi \pi \ell \nu$
nonresonant events.
We consider these dominant uncertainties in
more detail here.

\begin{table}
\begin{center}
\begin{tabular}{cccccc}
Systematic contribution  &  $\delta {\cal B}_{\rho}/{\cal B}_{\rho}$  & $\delta \vert V_{ub} \vert / \vert V_{
ub} \vert$ & $\delta \Gamma / \Gamma$  &
 $\delta \Gamma / \Gamma$  &
 $\delta \Gamma / \Gamma$  \\ \hline
$q^2$ range (${\rm GeV}^2/c^4$)  
& & &($0<q^2<7$) &($7<q^2<14$) & ($14<q^2<21$) \\ \hline
Simulation of detector               &  $\pm 9\%$ & $\pm 5\%$& $\pm 9\%$ & $\pm 6\%$ & $\pm 3\%$   \\
$b \rightarrow c \ell \nu$ composition               &  $\pm 2\%$ & $\pm 1\%$& $\pm 2\%$ & $\pm 6\%$ & $\pm 3\%$   \\
$b \rightarrow u \ell \nu$ composition             &  $\pm 6\%$ & $\pm 3\%$& $\pm 4\%$ & $\pm 9\%$ & $\pm 7\%$   \\
Integrated luminosity                                           &  $\pm 2\%$ & $\pm 1\%$& $\pm 2\%$ & $\pm 2\%$ & $\pm 2\%$     \\
Lepton identification                                &  $\pm 2\%$ & $\pm 1\%$& $\pm 2\%$ & $\pm 2\%$ & $\pm 2\%$    \\
Fake lepton rate                                     &  $\pm 1\%$ & $\pm 1\%$& $\pm 1\%$ & $\pm 3\%$ & $\pm 2\%$    \\
Fit technique                                        &  $\pm 5\%$ & $\pm 3\%$& $\pm 5\%$ & $\pm 5\%$ & $\pm 5\%$       \\ 
$f_{+-}/f_{00}$                                      &  $\pm 2\%$ & $\pm 1\%$& $\pm 3\%$ & $\pm 3\%$ & $\pm 3\%$   \\
$B \rightarrow \pi \pi \ell \nu$ nonresonant  &  $-8 \%$   & $- 4 \%$& $- 10\%$ & $-10 \%$ & $-10 \%$      \\
$\tau_B$                                             &  $\pm 1\%$ & $\pm 2\%$& $\pm 3\%$ & $\pm 3\%$ & $\pm 3 \%$   \\ \hline
Total systematic error                               &  $^{+13 \%}_{-15 \%}$ & $^{+7 \%}_{-8 \%}$&$^{+12 \%}_{-16 \%}$ &$^{+14 \%}_{-17 \%}$ &$^{+13 \%}_{-16 \%}$ \\   
(excluding model dep.) & & & & & \\ 
\end{tabular}
\end{center}
\caption{Summary of systematic errors.}
\label{table:sysbr}
\end{table}

The leptons in $b \rightarrow c \ell \nu$ background events
are primarily from $B \rightarrow D \ell \nu$, 
$B \rightarrow D^* \ell \nu$, and
$B \rightarrow J/ \psi X$ decays.  The $B \rightarrow D^{**}
\ell \nu$ contribution is small but is nevertheless included
in the fit.
For events with $\vert \Delta E \vert < 0.5$~GeV and
$\vert M(\pi \pi) -M(\rho) \vert
< 0.15$~GeV$/c^2$ for the $\pi \pi$ modes and 
$\vert M(\pi^+ \pi^- \pi^0) -0.782 \vert
< 0.05$~GeV$/c^2$ for the $\pi^+ \pi^- \pi^0$ 
mode, the Monte Carlo simulation predicts the following 
mix of $b \to c \ell \nu$ modes in HILEP (LOLEP): 
$23 \%$ ($17 \%$) for $B \to D \ell \nu$,
$48 \%$ ($76 \%$) for $B \to D^* \ell \nu$, 
$23 \%$ ($1 \%$) for $B \to J/\psi X$,
and
$5 \%$ ($5 \%$) for $B \to D^{**} \ell \nu$.

The $b \to c \ell \nu$
systematic error reflects the fit sensitivity to
the shape of the $\Delta E$ vs.~$M(\pi \pi)$ distribution 
determined from the
$b \rightarrow c \ell \nu$ Monte Carlo simulation.
(Recall that the fitting method already allows the data to
determine the overall normalization of the $b \rightarrow c \ell \nu$
background as well as the shape of its lepton-energy distribution.) 
To evaluate this systematic error, we vary the relative
sizes of the background sources.  For $B \rightarrow D \ell \nu$
and $B \rightarrow D^* \ell \nu$ variations are $\pm 20 \%$,
well beyond the current experimental uncertainties in their
branching fractions.
For $B \rightarrow D^{**} \ell \nu$ and $B \to J/\psi X$
we make variations of $\pm 40 \%$.
These uncertainties produce only a small effect in the
signal branching fraction, about $2 \%$.

The same method is used to evaluate the uncertainty due
to the $b \rightarrow u \ell \nu$ simulation.  
(The fit method allows the data to determine the
normalization of this background component in each 
lepton-energy bin.)  We
vary the relative contribution of each mode predicted by the ISGW2
model by $\pm 50 \%$.  The systematic error assigned is
the sum in quadrature of these variations.  

We also examine the
sensitivity of the fit to the  
$B \rightarrow \pi \ell \nu$ decay distributions.  
In contrast to $b \to u \ell \nu$, where the normalization
is allowed to vary in each lepton-energy bin of the
fit, the shape of the $B \to \pi \ell \nu$ lepton-energy
distribution is fixed by the Monte Carlo prediction.
We therefore vary the lepton-energy spectrum of the 
$B \to \pi \ell \nu$ Monte Carlo and
use form factors for $B \rightarrow \pi \ell \nu$
determined using several different methods (much as we have done for 
the $B \rightarrow \rho \ell \nu$ component of the fit).
In both cases, the $B \rightarrow \rho \ell \nu$ results are
robust against changes to the $B \rightarrow \pi \ell \nu$
Monte Carlo.  We include this uncertainty in the 
systematic error due to the $b \rightarrow u \ell \nu$ simulation.

The computation of $\vert {\bf p}_{\rm miss} \vert$
relies on the Monte Carlo simulation to adequately simulate the detector
response for all charged tracks and clusters in the event,
not just those used to form the $\rho+\ell$ candidate.
To evaluate our sensitivity to the details of our
detector Monte Carlo simulation, we vary the input
parameters of the Monte Carlo.  These variations include
conservative variations of the tracking efficiency, 
charged-track momentum resolution, CsI cluster identification 
efficiency, cluster energy resolution, and the simulation
of the detector endcaps.
In addition, we examine our sensitivity to the number of
$K^0_L$ mesons produced and the accompanying detector response.  

Finally, we assign a systematic error associated with a 
nonresonant 
$B \to \pi \pi \ell \nu$ contribution.  Like
signal events, nonresonant events
would have a $\Delta E$ distribution centered around
zero.  The $\pi \pi$ invariant-mass distribution, however,
would be somewhat different.  
Although we do not know how to describe the $M(\pi \pi)$ 
distribution of such a contribution, other properties help
us to distinguish resonant $B \to \rho \ell \nu$ from
nonresonant $B \to \pi \pi \ell \nu$.
The isospin ($I$) of the hadronic system in a 
$B \to \pi \pi \ell \nu$ decay must be either $I=0$
or $I=1$.  For $I=1$, where the relative production of
$\pi^{\pm} \pi^0$:$\pi^+ \pi^-$:$\pi^0 \pi^0$ is 2:1:0,
the relative orbital angular momentum ($L$)
must be odd.  The $L=1$ contribution is dominated 
by the $\rho$ resonance~\cite{lkgrev}.  Contributions
from $L=3,~5,~....$ are suppressed.  For $I=0$,
the relative production of $\pi^{\pm} \pi^0$:$\pi^+ \pi^-$:$\pi^0 \pi^0$ 
is 0:2:1, distinct from $I=1$.  An $I=0$
contribution will consist primarily of $L=0$.  In addition, the
$\pi^0 \pi^0$ mode, which has no resonant contribution,
is useful for constraining the size of
any nonresonant $B \to \pi \pi \ell \nu$ contribution.

Nonresonant events should also differ
from $B \to \rho \ell \nu$ events in their $q^2$ distribution.
We expect nonresonant events to occur mainly at
low $q^2$, where the daughter $u$ quark has a 
large momentum relative to the spectator quark.
As shown in Table~\ref{table:elepeffvsq2},
the efficiency of the $E_{\ell} > 2.3$~GeV requirement is 
highest at large $q^2$.  Thus, we expect to preferentially
select resonant $B \to \rho \ell \nu$ events.  

To assign a systematic error, we 
include a possible nonresonant contribution in our fit, using
two sets of assumptions for the $M(\pi \pi)$
and $q^2$ distribution.  First, we consider
a $M(\pi \pi)$ distribution with a broad Breit-Wigner shape having
$M=0.8$~GeV$/c^2$ and $\Gamma=1.0$~GeV$/c^2$.  
To simulate a $q^2$ distribution that is peaked at low
$q^2$, we
use $B \to \pi \ell \nu$ form factors from the ISGW2 model.
The second set of parameters
is designed such that the nonresonant contribution is
very similar to resonant $B \to \rho \ell \nu$ events.
We use 
$B \to \rho \ell \nu$ form factors (again from ISGW2) and
a $\rho$ Breit-Wigner for the parent $M(\pi \pi)$ 
distribution.  

Using these two sets of assumptions, we repeat our likelihood
fit with the additional freedom of a possible
nonresonant $B \to \pi \pi \ell \nu$ contribution.
As described above, we assume 
that this contribution
is primarily $I=0$.  A $20 \%$ $I=1$ contribution
accounts for a possible $L=3$ component.  Additionally, we
perform the fit with and without the $\pi^0 \pi^0$ mode.

In all
fits, we find a nonresonant component consistent with zero.  The
systematic error is assigned from the increase in the statistical
uncertainty on the $B \rightarrow \rho \ell \nu$ yield due to
the correlation with the nonresonant fit component.  This systematic
error is one-sided, as our nominal fit assumes no nonresonant
contribution.  The associated uncertainty is $8 \%$, which we
regard as conservative.

\section{${\cal B}(B \rightarrow \rho \ell \nu)$, 
$\vert V_{ \lowercase{ub}} \vert$, 
and $\Delta \Gamma$ results}
\label{sec:res}
\begin{figure}
\begin{center}
\epsfig{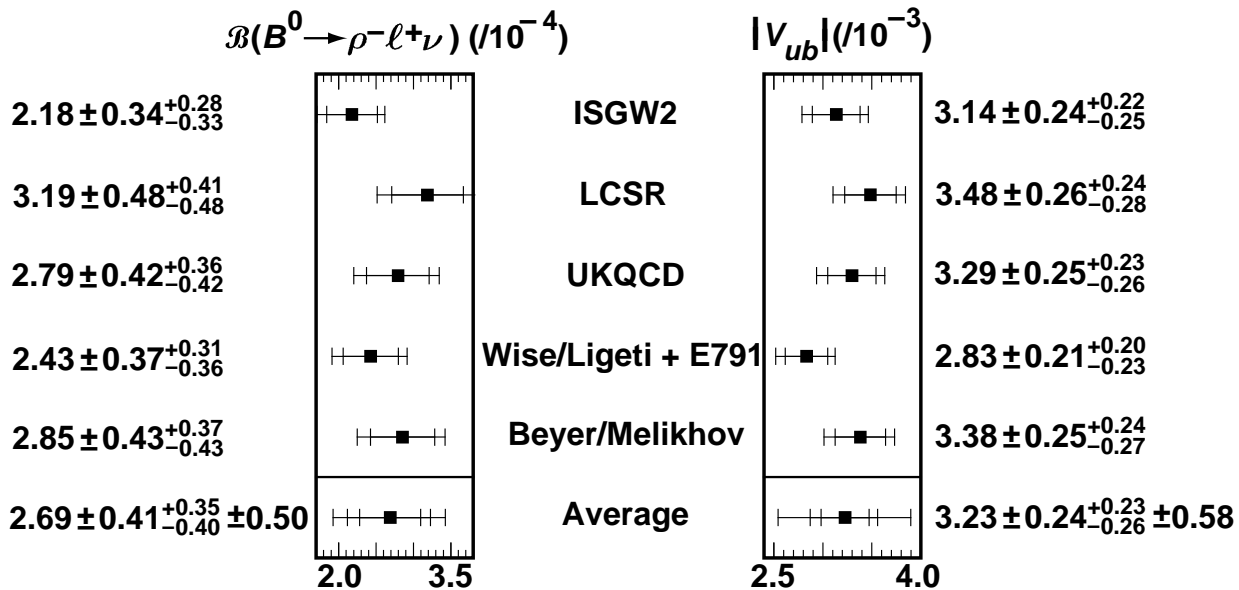}
\end{center}
\caption{${\cal B}(B^0 \rightarrow \rho^- \ell^+ \nu)$
 and $\vert V_{ub} \vert$ results.  
The errors are statistical,
systematic, and theoretical (on the averages), successively
combined in quadrature.  For the
branching fraction measurement, the theoretical error is
taken to be $1/2$ of the full spread of results.  For the $\vert V_{ub} \vert$
measurement there is an additional contribution
to the theoretical uncertainty due to the 
determination of $\tilde{\Gamma}_{{\rm thy}}$.} 
\label{fig:brres}
\end{figure}

To extract the 
$B \rightarrow \rho \ell \nu$ branching fraction
from our measured yields, we must extrapolate to the full
lepton-energy range.  We use form-factor 
models, as described in Section~\ref{sec:kine}, to determine
the signal efficiency.  The fitted yields themselves
depend only slightly on the 
set of form factors used.
The signal efficiency and yield for 
each form-factor model are presented in Table~\ref{table:sigyields}.
In addition, to extract
$\vert V_{ub} \vert$, we use the value of 
$\tilde{\Gamma}_{{\rm thy}}$ for each form-factor
model, along with $\tau_B$, to relate the branching fraction
result to $\vert V_{ub} \vert$, as given by
Eq.~\ref{eq:vub}.  Results for 
${\cal B}(B \to \rho \ell \nu)$ and $\vert V_{ub} \vert$
are presented in Fig.~\ref{fig:brres}.  

The final values for 
${\cal B}(B \to \rho \ell \nu)$ and $\vert V_{ub} \vert$ are  
the averages of the results obtained using 
the five form-factor models.  
We assign a systematic error to account for the substantial
spread in results among the form-factor models.
For the measurement of ${\cal B}(B \rightarrow \rho \ell \nu)$
this error is
assigned to be 1/2 the full spread among the five form-factor model results.
This uncertainty reflects our sensitivity to 
the different {\it shapes} of kinematic
distributions predicted by the models,
(mostly due to the $E_{\ell} > 2.3$~GeV acceptance.
For $\vert V_{ub} \vert$, we must also include an error due
to the uncertainty on the computation of
$\tilde{\Gamma}_{{\rm thy}}$, relecting
the different {\it normalizations} of the form-factor models.
The models quote errors on $\tilde{\Gamma}_{{\rm thy}}$ between
$17 \%$ and $50 \%$.  We have therefore assigned
a $30 \%$ error on $\tilde{\Gamma}_{{\rm thy}}$ 
(corresponding to a $15 \%$ error on $\vert V_{ub} \vert$),
rather than $15 \%$, which is the spread in predictions among models.
We find
\begin{eqnarray}
{\cal B}(B^0 \rightarrow \rho^- \ell^+ \nu) &=& 
(2.69 \pm 0.41^{+0.35}_{-0.40} \pm 0.50) \times 10^{-4}, \nonumber \\
\vert V_{ub} \vert &=& (3.23 \pm 0.24^{+0.23}_{-0.26} \pm 0.58) \times 10^{-3}.
\end{eqnarray}
\noindent The errors are statistical, systematic, and theoretical.
The dominant uncertainty on $\vert V_{ub} \vert$ arises from the
theoretical error on the normalization, $\tilde{\Gamma}_{{\rm thy}}$.
This $30 \%$ uncertainty (corresponding to a $15 \%$ error on $\vert V_{ub}
\vert$) 
is independent of the method used to measure ${\cal B}(B \to \rho \ell \nu)$,
and it is larger than the statistical error on 
$\vert V_{ub} \vert $ ($6 \%$) or the model dependence
of the detection efficiency ($19 \%$ on $\epsilon$ or
$9 \%$ on $\vert V_{ub} \vert$).

%%%%%%%%%%Put figure here

\begin{figure}
\begin{center}
\begin{turn}{-90}
\epsfig{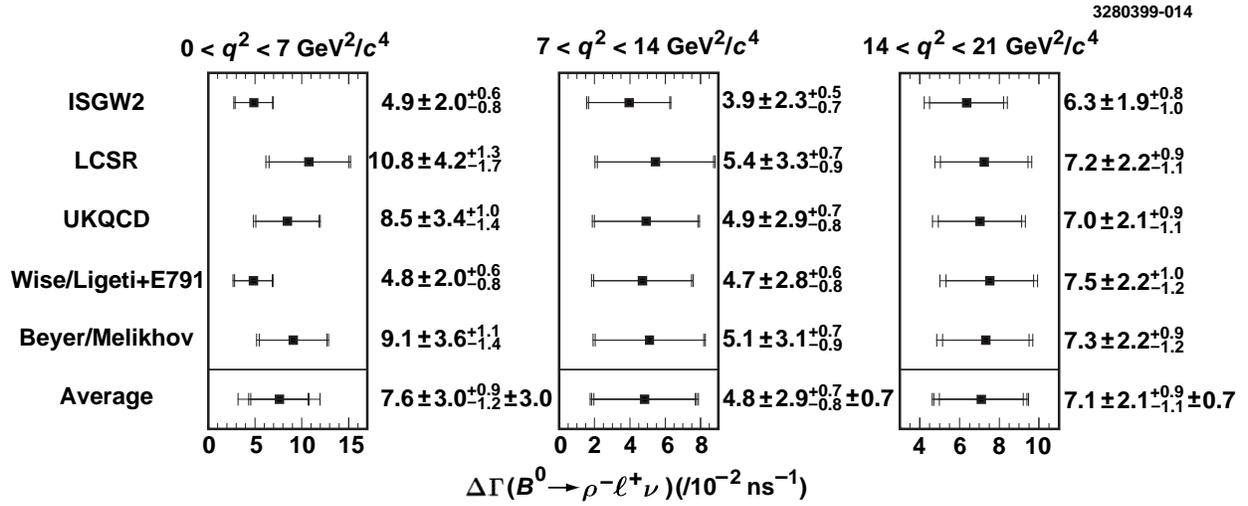}
\end{turn}
\end{center}
\caption{The partial width ($\Delta \Gamma$) in bins of $q^2$.  
The errors are statistical,
systematic, and theoretical (on the averages), 
successively combined in quadrature.
The theoretical error on each measurement is taken to be
$1/2$ of the full spread of results for the form-factor models.  }
\label{fig:q2res}
\end{figure}

\begin{figure}
\begin{center}
\begin{turn}{-90}
\epsfig{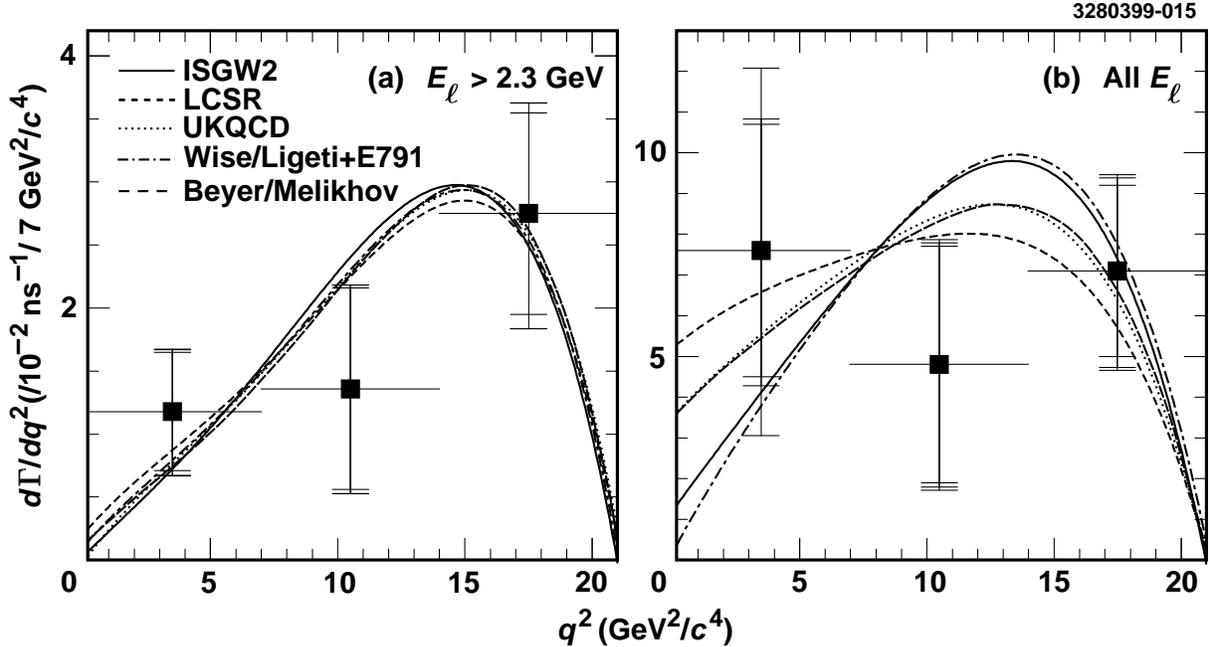}
\end{turn}
\end{center}
\caption{Comparison of measured $\Delta \Gamma$ distribution
(points with error bars)
with expectations from the form-factor models (curves) 
(a) for the $E_{\ell} > 2.3$~GeV
region and
(b) after 
the data have been extrapolated over the full $E_{\ell}$ range. 
The errors on the points are statistical,
systematic, and theoretical, successively combined
in quadrature.  Because the form-factor
models predict nearly the same $q^2$ distribution at
large lepton energy (a), we cannot distinguish between them.}
\label{fig:q2}
\end{figure}

Our results for $\Delta \Gamma $ in bins of $q^2$
are shown in Fig.~\ref{fig:q2res}.
As for the branching fraction and $\vert V_{ub} \vert$ results,
we compute an average over form-factor models and assign the theoretical
uncertainty to be one-half the full spread in results.  For these measurements,
the model dependence comes primarily from the variation in the efficiency
of the HILEP lepton-energy requirement.  We find
\begin{eqnarray}
\Delta \Gamma (0<q^2 < 7~{\rm GeV}^2/c^4) &=& 
(7.6 \pm 3.0 ^{+0.9}_{-1.2} \pm 3.0) 
\times 10^{-2} ~{\rm ns}^{-1}, \nonumber \\
\Delta \Gamma (7 < q^2 < 14~{\rm GeV}^2/c^4) &=& 
(4.8 \pm 2.9 ^{+0.7}_{-0.8} \pm 0.7) 
\times 10^{-2} ~{\rm ns}^{-1}, \nonumber \\
\Delta \Gamma (14<q^2 <21~{\rm GeV}^2/c^4) &=& 
(7.1 \pm 2.1 ^{+0.9}_{-1.1} \pm 0.6) 
\times 10^{-2} ~{\rm ns}^{-1},
\end{eqnarray}
\noindent where the errors are statistical, systematic, and theoretical.
Because the form-factor models predict
nearly the same $q^2$ distribution at large lepton energy,
we cannot distinguish among them, as 
shown in Fig.~\ref{fig:q2}.  The models do, however, agree
well with the $q^2$ distribution seen in the data.  At
high $q^2$, our lepton-energy requirement covers a
large fraction of the allowed lepton-energy range 
(as shown in Fig.~\ref{fig:dalitz}), and
we are able to measure a partial rate with a relatively
small theoretical error.

%%%%%%%%%%Put figure 8 here
%%%%%%%%%%Put figure 9 here

\begin{figure}
\begin{center}
\epsfig{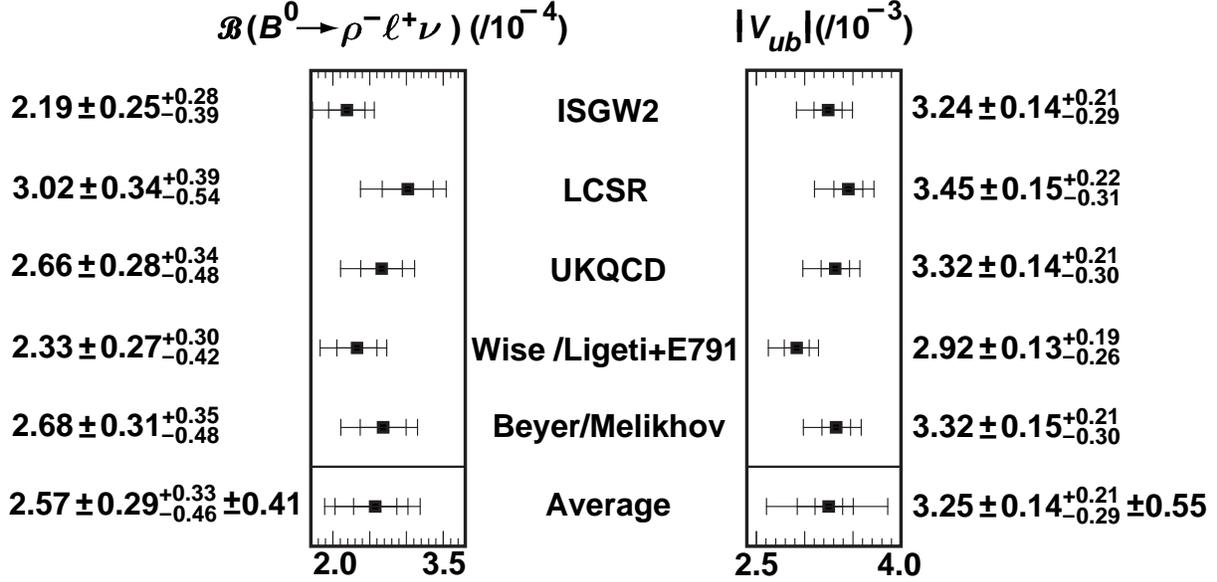}
\end{center}
\caption{Branching fraction and $\vert V_{ub} \vert$ results
averaged with the previous CLEO analysis.  
The errors are statistical,
systematic, and theoretical (on the averages), successively
combined in quadrature.  For the
branching fraction measurement, the theoretical error is
taken to be
$1/2$ of the full spread of results.  For the $\vert V_{ub} \vert$
measurement there is an additional contribution
to the theoretical uncertainty due to the 
determination of $\tilde{\Gamma}_{{\rm thy}}$.} 
\label{fig:brresave}
\end{figure}

Finally, we have formed an average for ${\cal B}
(B \rightarrow \rho \ell \nu)$ and $\vert V_{ub} \vert$
with the previously published CLEO exclusive $b \rightarrow u \ell \nu$
result~\cite{lkgprl}. 
The two methods share only a small fraction of events
and are 
therefore essentially statistically independent.
The published result has been updated
to consider the same set of form-factor models as used in this
paper.  (These results are described in Appendix~\ref{sec:newlkgresults}.)
A weight has been assigned to each analysis such as to
minimize the total error (statistical, systematic, and theoretical)
on the average.  Because Ref.~\cite{lkgprl} extracts
$\vert V_{ub} \vert$ using both $B \rightarrow \rho \ell \nu$
and $B \rightarrow \pi \ell \nu$ results,  we perform the average  
separately for the branching fraction and $\vert V_{ub} \vert$ measurements.
Figure~\ref{fig:brresave} shows the resulting averages
accounting for correlated systematic errors between the
two methods.  Averaging these results over form-factor
models, as described above, we obtain
 \begin{eqnarray}
{\cal B}(B^0 \rightarrow \rho^- \ell^+ \nu) &=& (2.57 \pm 0.29^{+0.33}_{-0.46}
\pm 0.41) \times 10^{-4}, \nonumber \\
\vert V_{ub} \vert &=& (3.25 \pm 0.14 ^{+0.21}_{-0.29} \pm 0.55) 
\times 10^{-3},
\end{eqnarray}
where the errors are statistical, systematic, and theoretical.

%%%%%%%%%%Put figure 10 here

\section{Conclusions and outlook}
\label{sec:conclusion}

We have performed a measurement of ${\cal B}(B \rightarrow \rho \ell \nu)$,
$\vert V_{ub} \vert$, and the $q^2$ distribution in 
$B \rightarrow \rho \ell \nu$ decay using a data sample of approximately
$3.3 \times 10^6$ $B \bar B$ pairs.  Using leptons
near the $b \rightarrow c \ell \nu$ lepton-energy endpoint,
we find
\begin{eqnarray}
{\cal B}(B^0 \rightarrow \rho^- \ell^+ \nu) &=& (2.69 \pm 0.41^{+0.35}_{-0.40}
\pm 0.50) \times 10^{-4}, \nonumber \\
\vert V_{ub} \vert &=& (3.23 \pm 0.24^{+0.23}_{-0.26} \pm 0.58) 
\times 10^{-3},  
\end{eqnarray}
where the errors are statistical, systematic, and
theoretical. The ${\cal B}(B \to \rho \ell \nu)$ result confirms
the previous CLEO measurement and has a comparable statistical 
precision.  This result is statistically
independent from the previous CLEO result
and has a somewhat smaller systematic error.
Averaging the measurements, we find
 \begin{eqnarray}
{\cal B}(B^0 \rightarrow \rho^- \ell^+ \nu) &=& (2.57 \pm 0.29^{+0.33}_{-0.46}
\pm 0.41) \times 10^{-4}, \nonumber \\
\vert V_{ub} \vert &=& (3.25 \pm 0.14 ^{+0.21}_{-0.29} \pm 0.55) 
\times 10^{-3}.
\end{eqnarray}
These values represent the current best 
CLEO results based on exclusive $b \to u \ell \nu$
measurements.  For the branching fraction, the
experimental ($^{+17 }_{-21} \%$) 
and theoretical uncertainties ($\pm 16 \%$) are comparable.
For $\vert V_{ub} \vert$, however,
the experimental 
uncertainties ($^{+8}_{-10} \%$) are substantially smaller than
the estimated theoretical error ($\pm 17\%$).  The
theoretical error on $\vert V_{ub} \vert$
contains a contribution from the uncertainty on the
detection efficiency ($\pm 8 \%$) and one from the overall
normalization ($\pm 15 \%$).

We have also measured the $q^2$ distribution 
in $B \to \rho \ell \nu$ decay in three 
bins.  We find
\begin{eqnarray}
\Delta \Gamma (0<q^2 < 7~{\rm GeV}^2/c^4) &=& 
(7.6 \pm 3.0 ^{+0.9}_{-1.2} \pm 3.0) 
\times 10^{-2} ~{\rm ns}^{-1}, \nonumber \\
\Delta \Gamma (7 < q^2 < 14~{\rm GeV}^2/c^4) &=& 
(4.8 \pm 2.9 ^{+0.7}_{-0.8} \pm 0.7) 
\times 10^{-2} ~{\rm ns}^{-1}, \nonumber \\
\Delta \Gamma (14<q^2 < 21~{\rm GeV}^2/c^4) &=& 
(7.1 \pm 2.1 ^{+0.9}_{-1.1} \pm 0.6) 
\times 10^{-2} ~{\rm ns}^{-1}.
\end{eqnarray}

Because we are sensitive to a 
large fraction of the allowed lepton-energy region
at high $q^2$, we 
measure the partial rate for $14<q^2 <21$~GeV$^2/c^4$ 
with a relatively
small theoretical uncertainty. This result is promising for future
analyses of $B \rightarrow \rho \ell \nu$ that use
the high lepton-energy region to determine $\vert V_{ub} \vert$,
if theoretical predictions
for $\Delta \tilde{\Gamma}_{{\rm thy}}$ at high $q^2$
can be made with good precision.  
For high lepton energies, the predicted shapes
of the $q^2$ distributions are virtually identical
for the different form-factor models.  The models,
therefore, cannot be distinguished on the basis
of measurements in this lepton-energy region alone.

To determine $\vert V_{ub} \vert$ more precisely using 
the full $q^2$ range, it
is important to include leptons whose energy is below the
lepton-energy endpoint region, so that 
the full $q^2$ distribution of $B \to \rho \ell \nu$ decay can
be measured.  Experimental results
on the $q^2$ distribution for leptons with $E_{\ell} < 2.0$~GeV
would help to improve form-factor models, and thus the 
measurement of $\vert V_{ub} \vert$. 

\section{Acknowledgements}
We gratefully acknowledge the effort of the CESR staff in providing us with
excellent luminosity and running conditions.
J.R. Patterson and I.P.J. Shipsey thank the NYI program of the NSF, 
M. Selen thanks the PFF program of the NSF, 
M. Selen and H. Yamamoto thank the OJI program of DOE, 
J.R. Patterson, K. Honscheid, M. Selen and V. Sharma 
thank the A.P. Sloan Foundation, 
M. Selen and V. Sharma thank Research Corporation, 
S. von Dombrowski thanks the Swiss National Science Foundation, 
and H. Schwarthoff and E. von Toerne thank 
the Alexander von Humboldt Stiftung for support.  
This work was supported by the National Science Foundation, the
U.S. Department of Energy, and the Natural Sciences and Engineering Research 
Council of Canada.

\appendix
\section{Update of previous CLEO results}
\label{sec:newlkgresults}

In this Appendix we present updated results of the previously
published CLEO analysis of 
$B \to \rho \ell \nu$ and $B \to \pi \ell \nu$,
which used $2.84 \times 10^6 B \bar B$ pairs~\cite{lkgprl}.  These
results are used in Section~\ref{sec:res}  to compute 
average values of ${\cal B}(B \to \rho \ell \nu)$ and 
$\vert V_{ub} \vert$.

We have updated these results
to use form-factor models 
for $B \to \rho \ell \nu$ and $B \to \pi \ell \nu$ decay
developed since Ref.~\cite{lkgprl}.  To compute 
${\cal B}(B \to \rho \ell \nu)$ and 
$\vert V_{ub} \vert$, a form-factor model is needed to
describe both $B \to \rho \ell \nu$ and $B \to \pi \ell \nu$
decay.  In general, we use predictions from the
same authors to describe both sets of 
form factors when both are predicted~\cite{isgw2,melikhov}.
For the fits using the UKQCD, LCSR, and 
Wise/Ligeti+E791 predictions to describe $B \to \rho \ell \nu$
events, we use the $B \to \pi \ell \nu$
predictions from Khodjamirian~{\it et al.}~\cite{ruckl} using
QCD sum rules (QCD SR). 

The fit results have been updated to use these form-factor
models as well as the $\tau_B$ results used in this
paper.  No other changes from Ref.~\cite{lkgprl}
have been made.  Results for
${\cal B}(B \to \rho \ell \nu)$,
${\cal B}(B \to \pi \ell \nu)$, 
and $\vert V_{ub} \vert$
are shown in Table~\ref{table:lkgres}.

\begin{table}
\begin{center}
\begin{tabular}{ccccc}
$\rho \ell \nu$ FF model & $\pi \ell \nu$ FF model & 
${\cal B}(\rho \ell \nu)$ ($/10^{-4}$) &
${\cal B}(\pi \ell \nu)$ ($/10^{-4}$) &
$\vert V_{ub} \vert$ ($/10^{-3}$) \\ \hline
ISGW2            &ISGW2         & 2.2 $\pm 0.4^{+0.4}_{-0.6}$ & 2.0 $\pm$ 0.5 $\pm 0.3$& 3.3 $\pm 0.2^{+0.3}_{-0.4}$ \\
LCSR             &QCD SR        & 2.8 $\pm 0.5^{+0.5}_{-0.8}$ & 1.6 $\pm$ 0.4 $\pm 0.3$& 3.4 $\pm 0.2^{+0.3}_{-0.4}$\\
UKQCD            &QCD SR        & 2.5 $\pm 0.5^{+0.5}_{-0.7}$ & 1.7 $\pm$ 0.4 $\pm 0.3$& 3.3 $\pm 0.2^{+0.3}_{-0.4}$\\
Wise/Ligeti+E791 &QCD SR        & 2.2 $\pm 0.4^{+0.4}_{-0.6}$ & 1.8 $\pm$ 0.4 $\pm 0.3$& 3.0 $\pm 0.2^{+0.3}_{-0.4}$\\
Beyer/Melikhov   &Beyer/Melikhov& 2.5 $\pm 0.5^{+0.5}_{-0.7}$ & 1.6 $\pm$ 0.4 $\pm 0.3$ & 3.3 $\pm 0.2^{+0.3}_{-0.4}$\\ 
\end{tabular}
\end{center}
\caption{Updated results of Ref.~\protect\cite{lkgprl}
for ${\cal B}(B^0 \to \rho^- \ell^+ \nu)$,
${\cal B}(B^0 \to \pi^- \ell^+ \nu)$, and
$\vert V_{ub} \vert$.  The errors are statistical and systematic.}
\label{table:lkgres}
\end{table}


\begin{thebibliography}{99}
\bibitem{vqq}
This notation applies to the decay of a charge -1/3 quark.  If
$Q$ is a +2/3 charge quark, $V_{q^{\prime}Q}$ should be 
replaced by $V^*_{Q q^{\prime}}$.
\end{thebibliography}
\end{document}